\documentclass{aa}
\newcommand{\figref}[1]{Fig.\,\ref{#1}}
\newcommand{\tabref}[1]{Table\,\ref{#1}}
\newcommand{\secref}[1]{Section\,\ref{#1}}
\usepackage{graphicx}
\usepackage{natbib}
\usepackage{txfonts}
\usepackage[]{hyperref}
\makeatletter
\renewcommand*\aa@pageof{, page \thepage{} of \pageref*{LastPage}}
\makeatother
\usepackage{xcolor}

\newcommand{\beq}{\begin{equation}}
\newcommand{\eeq}{\end{equation}}
\newcommand{\bea}{\begin{eqnarray}}
\newcommand{\eea}{\end{eqnarray}}

\begin{document}

\title{Disks in close binary stars}
\subtitle{$\gamma$-Cephei revisited}

\author{Lucas M. Jordan
	\inst{1},
	Wilhelm Kley
	\inst{1},
	Giovanni Picogna
	\inst{2}
	\and
	Francesco Marzari
	\inst{3}
}

\institute{$^1$ Institut für Astronomie und Astrophysik, Universität Tübingen,
	Auf der Morgenstelle 10, 72076 Tübingen, Germany\\
	\email{lucas.jordan@uni-tuebingen.de}\\$^2$ Universit\"ats-Sternwarte, Fakult\"at f\"ur Physik, Ludwig-Maximilians-Universit\"at M\"unchen, Scheinerstr 1,
	D-81679 M\"unchen, Germany\\$^3$ Dipartimento di Fisica, University of
	Padova, via Marzolo 8, 35131 Padova, Italy
}
\date{Received 5 May 2021; Accepted 27 July 2021}
\abstract
{	Close binaries ($a_\mathrm{bin} \leq 20\,\mathrm{au}$) are known to harbor
	planets, yet planet formation is unlikely to succeed in such systems.
	Studying the dynamics of disks in close binaries can help to understand how
	those planets could have formed.}
{ We study the impact that numerical and physical parameters have on the
dynamics of disks in close binaries. We use the $\gamma$-Cephei system as an
example and focus on disk quantities such as disk eccentricity and the
precession rate as indicators for the dynamical state of the disks.}
{ We simulate disks in close binaries by performing two-dimensional radiative
  hydrodynamical simulations using a modified version of the \textsc{Fargo}
  code. First, we perform a parameter study for different numerical parameters
  to confirm that our results are robust. In the second part, we study the
  effects of different masses and different viscosities on the disks' dynamics.}
{ Previous studies on radiative disks in close binaries used too low resolutions
  and too small simulation domains, which impacted the disk's dynamics. We find
  that radiative disks in close binaries, after an initialization phase, become
  eccentric with mean eccentricities between $0.06$ to $0.27$ and display a slow
  retrograde precession with periods ranging from $4 - 40T_\mathrm{bin}$ which
  depends quadratically on the disk's mean aspect ratio. In general, the disks
  show a coherent, rigid precession which can be broken, however, by changes in
  the opacity law reducing the overall eccentricity of the disk.}
{}


\maketitle
%

\section{Introduction}
Up to now, about a dozen known planets have been detected that reside inside a
close binary system ($a_\mathrm{bin} \lessapprox 20\,\mathrm{au}$) orbiting one
of the stars in the system, in an S-type orbit \citep{thebault2015planet}. One
well-studied example is the $\gamma$-Cephei system, which is a close binary
system with a semi-major axis of $20\,\mathrm{au}$ and eccentricity of $0.4$
\citep{Endl2011NewsFT}. It harbors a giant planetary companion
\citep{Hatzes-2003ApJ...599.1383H} with a mass of $m\,sin\,i = 1.85\,M_{Jup}$
distanced $2\,\mathrm{au}$ from the more massive companion
\citep{Endl2011NewsFT}. This minimum planet mass was determined via radial
velocity measurements, but recently \citet{benedict2018mass} claimed a much
larger actual mass of $\sim9.4\,M_{Jup}$ and a non-coplanar system using Hubble
Space Telescope astrometry data. The origin of planets in these close binaries
incites many questions, as previous studies have shown that the standard planet
formation process is highly problematic at the observed positions of the planets
due to the presence of a close stellar companion (see
\citet{thebault2015planet,Thebault2019Planets} for a full discussion).
\linebreak \indent The position of the planet in $\gamma$-Cephei is not too far
from the orbital stability limit of $\sim4\,\mathrm{au}$ \citep{holman1999long,
pichardo2005circumstellar}. At this position, the companion plays an important
role in the planet formation process. The planetesimal accretion phase was found
to be the phase that is most sensitive to gravitational disturbances. The
eccentricity of the planetesimals is excited by the companion star and damped by
the gas disk. The size dependency of the gas drag causes misalignment in the
pericenters for different-sized planetesimals. This leads to orbit-crossings
that cause destructive collisions and prevent further growth
\citep{thebault2006relative, thebault2008planet}. The relative velocities
between the planetesimals become even larger if the hosting disk is eccentric
\citep{paardekooper2008planetesimal}. Later, \citet{rafikov2013planet} noted
that the disk gravity has a dominant effect on the planetesimal's dynamics and
concluded in \citet{rafikov2014planet} that planetesimal growth could succeed in
close binaries given that the disk is apsidally aligned to the binary and has a
very small eccentricity ($e_\mathrm{d} \leq 0.02$). Therefore, precise modeling
of disks in close binaries is necessary to understand how planet formation can
succeed.

Within the context of superhumps in cataclysmic variables
\citep{Pearson-2006MNRAS.371..235P}, the importance of tidal effects by the
secondary star was demonstrated in the simulations by
\citet{1988MNRAS.232...35W}. Later, eccentric disks in circular close binaries
were studied by \citet{lubow1991model} who found, using linear analysis, that
the $m=3$ eccentric inner Lindblad resonance can drive the growth of disk
eccentricity in such systems. This was also confirmed numerically in
\citet{lubow1991simulations} and \citet{artymowicz1994dynamics}. Though later it
was shown that the $m = 3$ resonance is not the only cause for eccentricity
growth in disks in close binaries either by removing the resonance from the
potential or by using a larger mass ratio of the binary that truncates the disk
to sizes such that the $m = 3$ resonance is not contained in the disk anymore
\citep{kley2008simulations,marzari2009eccentricity}.
Generally, hydrodynamical simulations have shown that in close binary star
systems, either circular or eccentric, the disks regularly develop large
eccentricities $e_\mathrm{d} \geq 0.2$ and start precessing
\citep{paardekooper2008planetesimal,kley2008simulations,
marzari2009eccentricity,marzari2012eccentricity,muller2012circumstellar}. In
those simulations, it has been seen that the size of the disk is an important
indicator for the final disk eccentricity. Disks that are more exposed to the
perturbations of the companion tend to develop larger eccentricities. The size
of the disk is determined by the viscosity of the gas with higher viscosity
leading to larger disks, and by the tidal effect of the secondary star that
leads to disk truncation.
 
Concerning numerical considerations, \citet{paardekooper2008planetesimal} noted
that the numerical methods used in the simulations can drastically affect the
dynamics of the disk. They found that their simulations, which used the RODEO
code \citep{Paardekooper-2006A&A...450.1203P} with the diffusive minmod limiter,
produced disks in a quiet state that have low eccentricity and show no
precession. This quiet state was not affected by changes to resolution or
boundary conditions. When the less diffusive superbee flux limiter was used, the
disks entered an excited state with precession and high eccentricity that
strongly depended on the resolution and boundary condition used. Disks in this
low eccentric state have also been observed in locally isothermal simulations
that include self-gravity \citep{marzari2009eccentricity}, or in radiative
simulations \citep{muller2012circumstellar, marzari2012eccentricity}, or SPH
simulations \citep{martin2020evolution}.

After all, it is currently not clear how numerical considerations and physical
parameters impact the dynamics of circumstellar disks in close binary stars. To
better understand the process of planet formation in binary stars and gain more
insight into the subgroup of superhump systems in cataclysmic variables, further
development in theoretical and numerical models of disk dynamics in binary star
systems is warranted. Additionally, there have been massive advances in
computational resources since those studies were performed, and we decided to
revisit the problem of disks in close binaries and continue the studies
presented in \citet{muller2012circumstellar}. The focus of this paper lies again
on the system $\gamma$-Cephei, being one of the outstanding sample systems, with
the first planet discovered in a close binary star
\citep{Hatzes-2003ApJ...599.1383H}. In our study, we treat physically realistic
disks by including internal heating and radiative cooling.

This paper is structured as follows. In \secref{sec:model} we present our model
used for the simulations and the methods for analysis.  We briefly explain our
standard model in \secref{sec:standard}. Tests of different grid resolutions,
domain-sizes and boundary condition are presented in \secref{sec:convergence},
and the results from simulations with different disk masses and viscosities are
shown in \secref{sec:physical}. We discuss global trends of the disks in detail
in \secref{sec:ecc_prec}, highlight difficulties when simulating disks in close
binaries in \secref{sec:discussion}, and finally summarize our results in
\secref{sec:summary}.

\section{Model}
\label{sec:model}
In this work, we simulated two-dimensional, radiative disks, using a modified
version of the \textsc{Fargo} code including the energy equation. The code
solves the vertically integrated hydrodynamical equations on a polar coordinate
system ($r - \phi$), centered on the primary star. It uses an upwind scheme with
a second-order flux limiter \citep{van1977towards}, linear interpolation to
reconstruct variables, and the fast advection in rotating gaseous objects
(FARGO) method for the azimuthal advection, for details, see
\citet{masset2000fargo}. For shock smoothing, artificial viscosity is used as
described in \citet{stone1992zeus1}. The simulated disks are
non-self-gravitating and the gravitational back-reaction from the disks onto the
binary stars is neglected. The orbit of the binary is integrated with a
fifth-order Runge-Kutta method. The source terms for the energy equation include
viscous heating and radiative cooling alongside compressive heating. For the
Rosseland mean opacity, we used the piece-wise power-laws as put forward in
\citet{lin1985dynamical}. The full equations and physical assumptions are
specified in detail in \citet{muller2012circumstellar}, and we do refer to that
work for more information.

\begin{table}[!t]
	\centering
	\caption{Physical and numerical parameters for initializing the standard model.\label{table:parameters}}
\renewcommand{\arraystretch}{1.2}
\begin{tabular}{l l}
	\hline\hline
	Primary star mass	&	$1.4\,\mathrm{M}_{\sun}$	\\
	Secondary star mass	&	$0.4\,\mathrm{M}_{\sun}$	\\
	Semi-major axis		&	$20\,\mathrm{au}$			\\
	Eccentricity		&	0.4							\\
	Orbital period		&	$66.7\,\mathrm{a}$			\\
	\hline
	Disk mass			&	$0.01\,\mathrm{M}_{\sun}$	\\
	$\alpha$ viscosity	&	$10^{-3}$					\\
	Adiabatic index		&	$7/5$						\\
	\hline
	Initial Density profile	&	$\propto	r^{-1}$		\\
	Initial Temperature profile	&	$\propto	r^{-1}$	\\
	\hline
	$\mathrm{R}_{\mathrm{min}} - \mathrm{R}_{\mathrm{max}}$	&	$0.2\,-\,12\,\mathrm{au}$\\
	Grid ($\mathrm{N} _{\mathrm{r}} \times \mathrm{N}_{\mathrm{\varphi}}$)	&
	$760	\times	1162$\\
	\hline
\end{tabular}
\tablefoot{Since there is no gravitational feedback from the disk onto the
	stars, the binary orbital parameters do not change during the simulation.
	The disk mass of $10^{-2} M_\odot$ equates to $\Sigma(1\,\mathrm{au}) =
	2570\,\mathrm{g}\,\mathrm{cm}^{-2}$.}
\end{table}

\subsection{Initialization}
The physical and numerical parameters of our fiducial model are presented in
Table\,\ref{table:parameters}. For the initial temperature, we use a power-law
with $T(r) \propto r^{-1}$, which leads to a disk with constant aspect ratio,
$h$, see Eq.\,\eqref{eq:aspect-ratio}. The reference temperature is chosen such
that $h = 0.05$. Since viscous heating and radiative cooling are explicitly
included in the models, the actual disk temperature will change quickly during
the simulations. For the initial density, we chose the same power-law $\Sigma(r)
\propto r^{-1}$. Because the disk around the primary is truncated by the tidal
forces of the secondary, the initial disk surface density is smoothly reduced
beyond $6\,\mathrm{au}$ to start the disk closer to the equilibrium size. The
total initial disk mass of our fiducial model (including this tapering) is
$10^{-2} M_\odot$ which equates to $\Sigma(1\,\mathrm{au}) =
2570\,\mathrm{g}\,\mathrm{cm}^{-2}$. The initial radial velocity is set to zero,
and the initial angular velocity is set to the Keplerian angular velocity.

\begin{figure*}[!th]
	\centering
	\includegraphics[width=180mm]{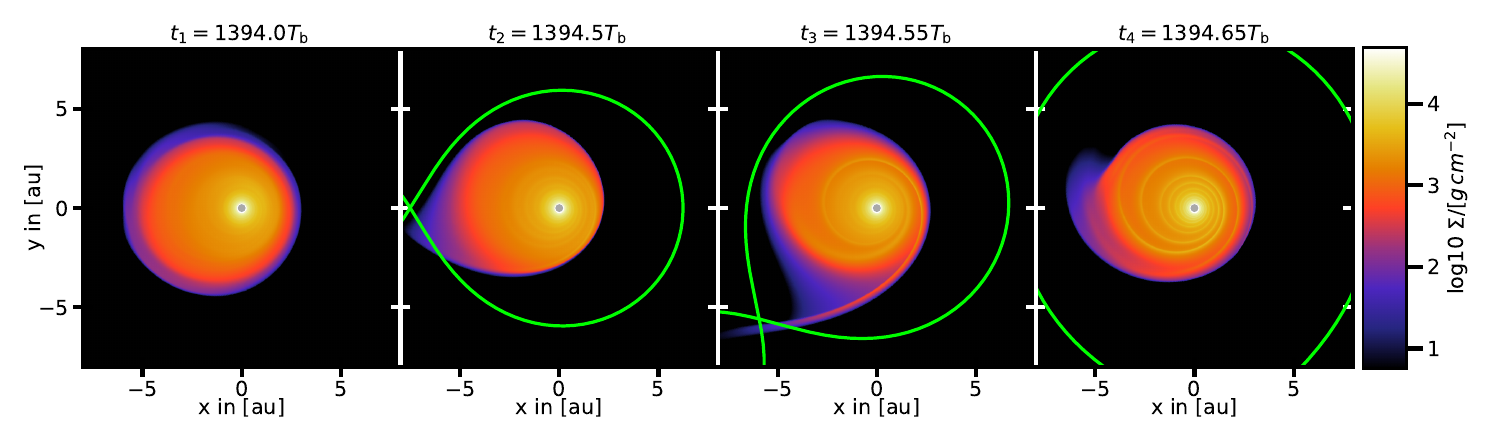}
	\caption{\label{fig:timeline} Disk surface density during one binary orbit
	for the standard parameters, see \tabref{table:parameters}, after the
	equilibrium has been reached. The green line indicates the Roche lobes of
	the stars. In the leftmost image ($t_1$) the binary is at apastron and the
	disk is unperturbed. At the binary's periastron ($t_2$) the disk is
	maximally distorted and has the highest eccentricity. After the periastron
	passage ($t_3$) two spiral arms are launched that wind inward and dissipate
	again ($t_4$). Most of the disk's mass loss happens in the time frame
	between $t_2$ and $t_4$ that spans 15\,\% of the binary's period.}
\end{figure*}
\subsection{Boundary conditions}
We use outflow boundaries at the outer edge of the computational domain for all
our simulations. At the inner edge of the domain, reflective boundaries are
used, unless stated otherwise. The implementation of the boundaries is identical
to \citet{muller2012circumstellar} (it is important to note that the reflective
boundary condition in their section 2.2 is missing a minus sign, and it should
read $v_{0j} = -v_{2j}$, for the radial velocity). We added wave damping zones
at the inner edge to prevent waves from being reflected into the domain by
damping the radial velocity $v_r$ to zero and the surface density $\Sigma$ and
internal energy $e$ to the azimuthal average by using the damping mechanism as
described by
\citet{de2006comparative}
\begin{equation}
  \frac{\mathrm{d}x}{\mathrm{d}t} = \frac{x - x_0}{\tau}f(r) \,,
\end{equation}
where $x \in \{v_r, \Sigma, e\}$, $x_0$ is the desired value, $\tau$ is the
damping timescale and $f(r)$ is a quadratic ramp-up function that rises from 0
to 1 from the start to the end of the damping zone. The wave damping zone ranges
from $1.1 R_\mathrm{min}$ to $R_\mathrm{min}$. For the damping timescale we used
$\tau = 10^{-3} 2\pi \Omega_\mathrm{k}(R_\mathrm{min})^{-1}$ where
$\Omega_\mathrm{k}$ is the Keplerian angular velocity around the primary. We
also tested slower wave damping and no wave damping and found that it has little
effect on the disk's dynamics.
 
\subsection{Numerical considerations}
To avoid numerical instabilities from too low density or energy values at the
outer border of the disk, we used a density floor of $\Sigma_\mathrm{floor} =
10^{-7} \cdot \Sigma_0$ where $\Sigma_0 = \Sigma(1\,\mathrm{au})|_{t=0}$ and a
temperature floor of $T_\mathrm{floor} = 10\,\mathrm{K}$. 
We use the $\alpha$-prescription by \citet{shakura1973alpha} to model the
viscosity and name simulations for our viscosity study by their $\alpha$ value.
Inside plots, we compare the simulations by their kinematic viscosity, as it is
the relevant quantity that affects the disk's dynamics. The coefficient of the
kinematic viscosity is defined as
\begin{equation}
\nu = \alpha
c_s^2\Omega_\mathrm{k}^{-1} \,,
\end{equation}
where $c_s$ is the adiabatic sound speed.

We also tested the disk dynamics of our standard case using additionally the
 \textsc{Pluto} code \citep{mignone2007pluto}. These results, presented in
 Appendix\,\ref{sec:codes}, showed similar behavior.
\begin{figure}[!t]
	\centering
	\includegraphics[width=88mm]{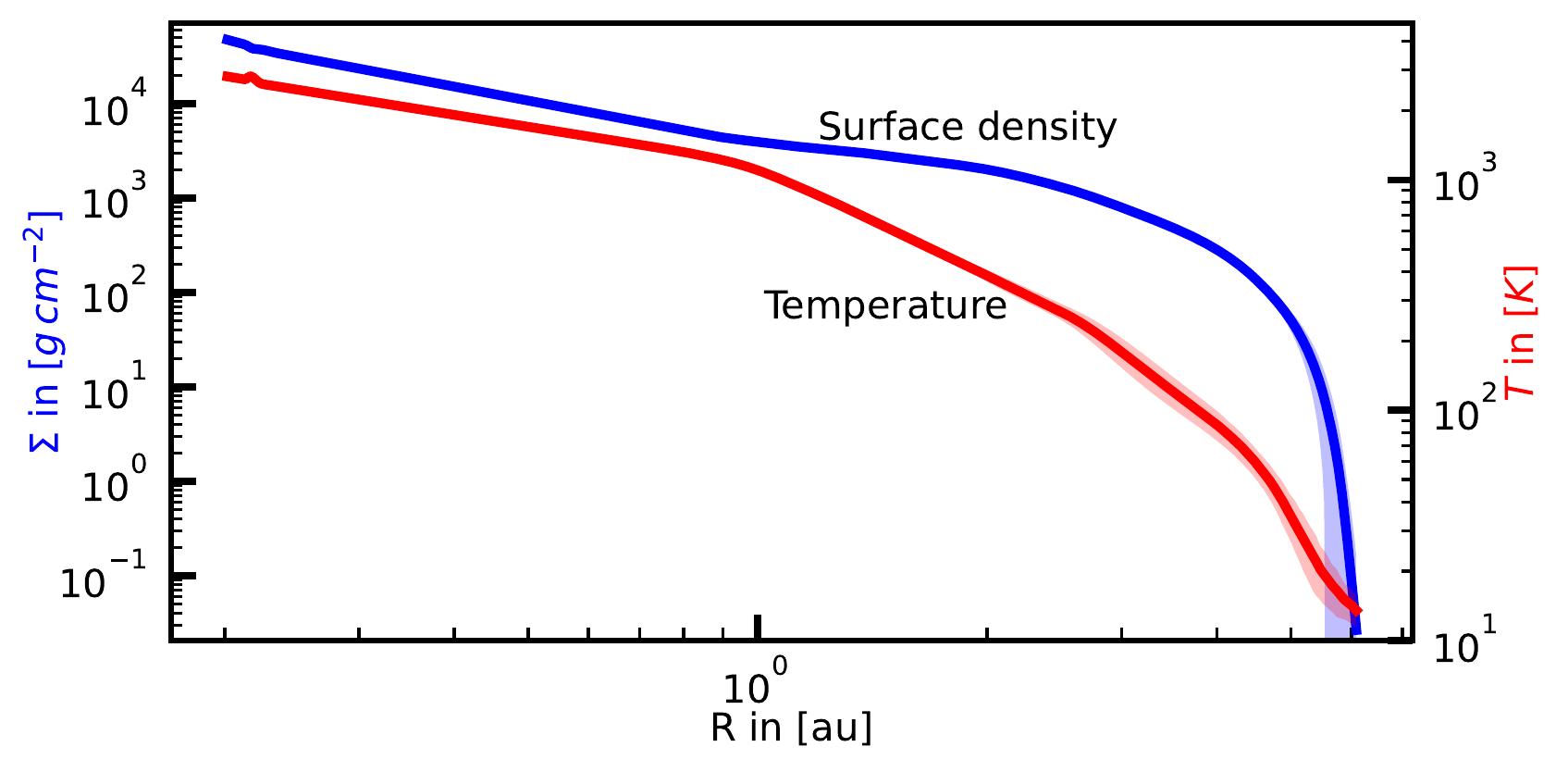}
	\caption{Time-averaged radial profile of the surface density and gas
	temperature of our fiducial model. Solid lines are averaged over 200
	snapshots taken at the binary apastron during the simulation time from
	$1300\,T_\mathrm{bin}$ to $1500\,T_\mathrm{bin}$. The shaded areas show the
	$1\sigma$ variations.  }
	\label{fig:fiducial_model}
\end{figure}
\begin{figure}[!th]
	\centering
	\includegraphics[width=88mm]{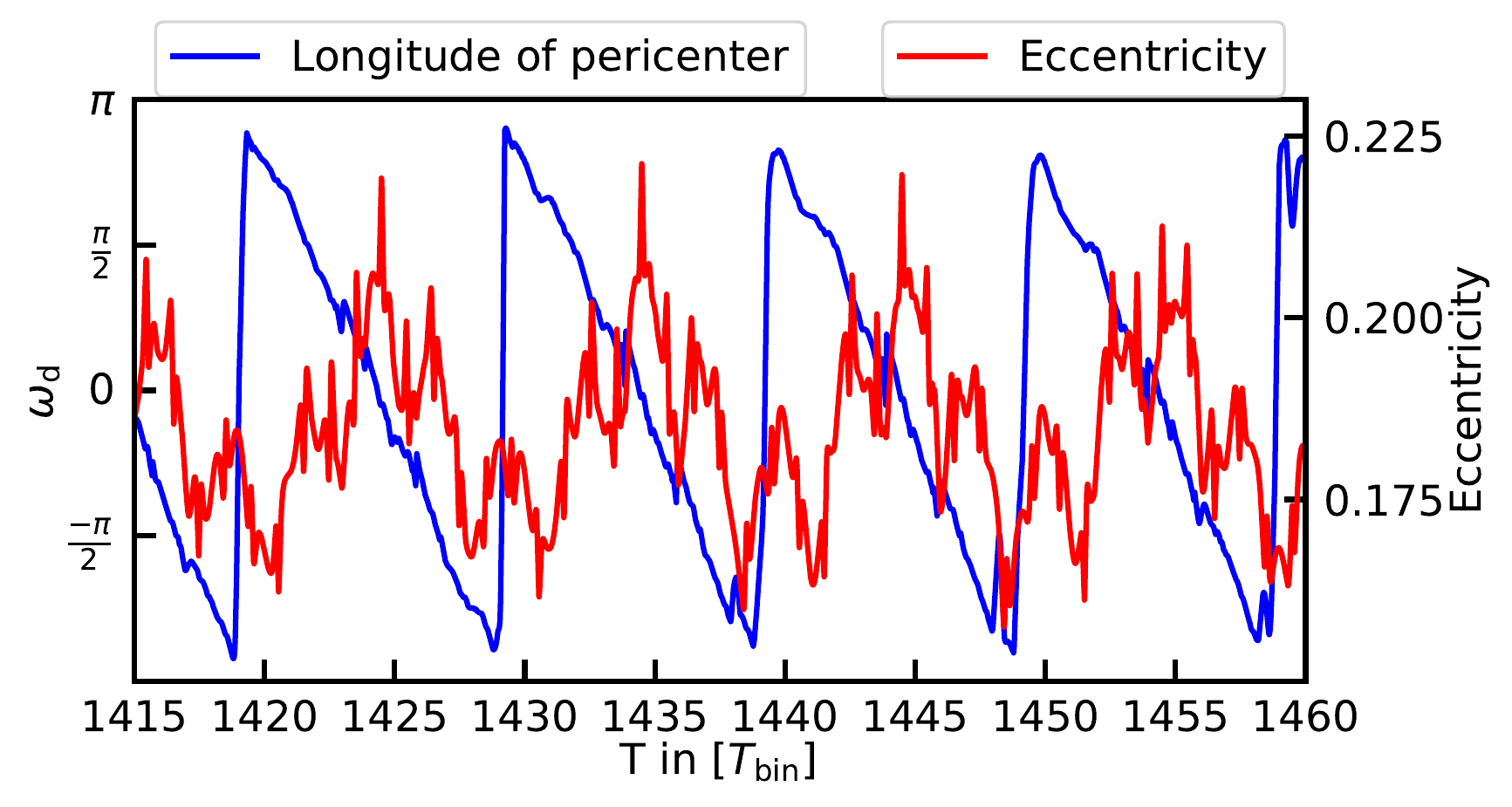}
	\caption{Time evolution of the mass-weighted disk eccentricity and the mass
	weighted longitude of pericenter for the fiducial model. The disk has the
	lowest eccentricity and has the least variation during one binary orbit when
	the longitude of pericenter is aligned with the binary periastron
	($\omega_\mathrm{d} = \pm\pi$). From that point the disk's eccentricity and
	variation during one binary orbit increases and reaches a maximum when the
	longitude of pericenter is aligned with the binary apastron
	($\omega_\mathrm{d} = 0$).}
	\label{fig:peri_ecc}
\end{figure}

\subsection{Analysis}
We describe the disk's dynamics by its global eccentricity $e_\mathrm{d}$ and
longitude of pericenter, $\omega_\mathrm{d}$. These are calculated from the
mass-weighted average over all cells as has been used in previous studies
\citep{kley2008simulations}. The desired quantity is calculated for each cell by
assuming that the gas parcel moves on a 2-body Keplerian orbit around the
primary star, then the average over the whole disk is calculated as
\beq
\label{eq:global-average}
   f_\mathrm{d} = \frac{\iint f_\mathrm{d} (r,\varphi) \Sigma(r,\varphi)
   r\mathrm{d}r\mathrm{d}\varphi}{M_\mathrm{Disk}} \,,
\eeq
where $f_\mathrm{d}  \in \{ e_\mathrm{d} , \omega_\mathrm{d} , h , \nu \}$.
Similarly, the radial profile of a disk quantity is calculated as 
\beq
\label{eq:phi-average}
f_\mathrm{d}(r) = \frac{\int f_\mathrm{d}(r,\varphi) \Sigma(r,\varphi)
\, \mathrm{d}\varphi}{\int \Sigma(r,\varphi) \, \mathrm{d}\varphi} \,.
\eeq
In the presentation of our results we refer to the disk temperature mostly
through the aspect ratio, $h$, which is the ratio of the vertical pressure scale
height of the disk, $H$, over the radius. It is given by
\beq
\label{eq:aspect-ratio}
     h = \frac{H}{r}  = \frac{c_\mathrm{s,iso}}{r \Omega_\mathrm{K}} \,,
\eeq
where the second equality follows for standard thin accretion disks.

Before engaging in our parameter studies, we first checked if resolution and
domain size is sufficient to resolve the disk dynamics. These studies are
presented in \secref{sec:convergence} below. Afterward, we varied the mass of
the disk between $2\cdot10^{-3}\, M_{\odot}$ to $5\cdot10^{-2}\, M_{\odot}$ the
$\alpha$-viscosity parameter from $10^{-4}$ to $10^{-2}$, and document the
effects that these variations have on the disks' dynamics.

\section{Standard model}
\label{sec:standard}
We present the physical and numerical parameters of our fiducial model in
Table\,\ref{table:parameters}. In deviation from the standard model used in
\cite{muller2012circumstellar}, which was also based on the $\gamma$-Cephei
system, we reduced the viscosity parameter from $\alpha = 10^{-2}$ to $\alpha =
10^{-3}$, because this better represents the average viscosity found in recent
disk observations
\citep[e.g.,][]{rafikov2017protoplanetary,sellek2020dusty,trapman2020observed}.
Additionally, we increased the numerical resolution of the grid and the extent
of the simulation domain.

The general behavior of the disk during a binary revolution is displayed in
\figref{fig:timeline}. During the binary apastron, the disk is eccentric but the
overall density is smooth, without any spiral features. Just before periapse,
the disk becomes visibly affected by the companion and is extended toward the
secondary in a shape that follows the Roche lobe of the primary. Mass loss
occurs shortly after the periapse, during the same time, two spiral arms develop
that wind toward the disk's center and dissipate. 

The radial disk structure is displayed in \figref{fig:fiducial_model} using the
time-averaged radial surface density and temperature profiles. The transition
between opacity power-laws are visible as bends in the temperature profile, such
as the bend at $1\,\mathrm{au}$ caused by the sublimation of dust at
$1100\,\mathrm{K}$. Inside the dust sublimation region, the temperature gradient
becomes shallower due to the lower opacity, and the surface density gradient
becomes steeper due to the disk being in a viscous equilibrium. Beyond
$4.5\,\mathrm{au}$, surface density drops rapidly because of the tidal
truncation of the binary. The kink in temperature close to the inner edge is
caused by the wave damping region.

All of our simulations produced retrograde precessing disks with one precession
period, lasting between $4 - 40T_\mathrm{bin}$, depending on their temperature.
The effects of the precession on the disk eccentricity are shown for the
standard model in \figref{fig:peri_ecc}. When the disk's apocenter is aligned
with the binary's periastron, the disk is most affected by the binary's
periastron passage, resulting in the largest perturbations (seen as the largest
spikes in \figref{fig:peri_ecc}). At this point, the disk's eccentricity is also
the largest. When the disk's longitude of pericenter is moving toward the
binary's periastron, it becomes less affected by the companion, resulting in
weaker perturbations during the binary's periastron passage and a waning disk
eccentricity.
\section{Numerical convergence}
\label{sec:convergence}
Before we present our studies on the influence of the physical parameter, we
first showcase in this section the impact of the numerical parameters on the
disk's behavior in a close binary star. Starting from our standard model in
\tabref{table:parameters}, we varied one numerical parameter at a time and
studied their effect on the dynamical evolution on the disk. A suitable
indicator for the dynamical state of the disk is the mass averaged global disk
eccentricity, $e_\mathrm{d}$, calculated according to
Eq.\,\eqref{eq:global-average}, and we present our finding in terms of
$e_\mathrm{d}$.
\subsection{Grid resolution}
\begin{figure}[!t]
	\centering
	\includegraphics[width=88mm]{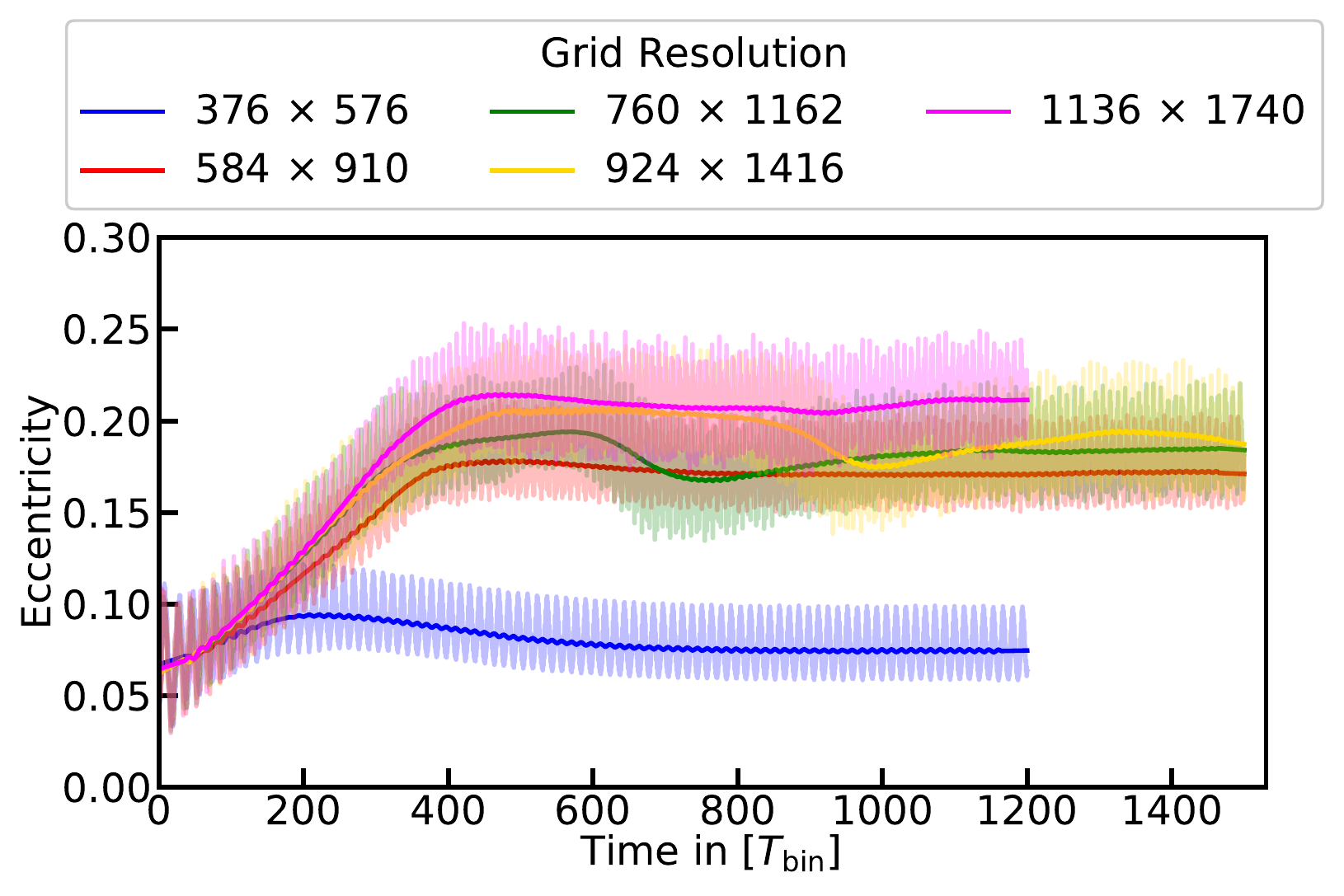}
	\caption{Time evolution of the mass-weighted disk eccentricity (see
	Eq.\,\eqref{eq:global-average}) for different resolutions for a logarithmic
	grid. Solid lines are time averaged values while the transparent lines are
	the simulation data.}
	\label{fig:log_resolution_ecc}
\end{figure}

In \figref{fig:log_resolution_ecc} the disk's eccentricity is shown for
different resolutions of the logarithmic grid. For our standard model (green
line in \figref{fig:log_resolution_ecc}) it takes the disk around 400 orbits to
reach an equilibrium state (similar values were found in
\cite{kley2008simulations} and \cite{muller2012circumstellar} for locally
isothermal simulations). The final eccentricity reached in this equilibrium
state depends on the numerical resolution. A too low resolution can damp the
disk's eccentricity significantly, as seen clearly in the lowest resolution
case, $376\times576$, in \figref{fig:log_resolution_ecc}, which ends up in a
different, low eccentric state compared to higher resolution cases. In
simulations with higher resolutions (with more than six cells per radial
pressure scale height) $e_\mathrm{d}$ grows until a new equilibrium state is
reached with $e_\mathrm{d} \approx 0.17 - 0.22$ for the various resolutions. Our
fiducial model has a resolution of eight cells per radial pressure scale height
to leave enough headroom to conduct simulations with colder disks (and therefore
fewer cells per scale height) on the same grid. Other works, which conducted
numerical convergence studies, also found that resolutions of around eight cells
per scale height are required to reach convergence, for example,
\citet{thun2017numerical} and \citet{oliva2020fragmentation}. In the high
eccentric state, the disks show a retrograde precession with a period of $(9.5 -
10.2)\,T_\mathrm{bin}$, which is always identical to the corresponding
oscillation period of $e_\mathrm{d}$, as shown in \figref{fig:peri_ecc} for the
standard case.
\linebreak \indent We repeated the resolution test with an arithmetic grid and
found the same behavior with comparable final disk eccentricity. Although, a
significantly higher resolution of $1846\times1800$ on an arithmetic grid was
required to match the simulation using a $760\times1162$ resolution on a
logarithmic grid. This is caused by the fact that the arithmetic grid has a
lower resolution (cells per pressure scale height) in the inner disk region and
instead has a higher resolution at the periphery than the logarithmic grid.
Reaching convergence between the two grids indicates that our disk is
sufficiently resolved throughout the whole domain and that the numerical
viscosity should be negligible.
\linebreak \indent We take these resolution studies as a sign that the excited,
eccentric disk state is physical and that the low eccentric state at lower grid
resolutions is only numerical. Since increasing the resolution further above our
standard model, has little effect on the dynamics, and for computational
reasons, we choose the $760\times1162$ resolution as our standard. Considering
our new results, it turns out that previous studies of radiative disks
\citep[e.g., ][]{muller2012circumstellar}  
may have suffered from too low grid resolution.
\begin{figure}[!t]
	\centering
	\includegraphics[width=88mm]{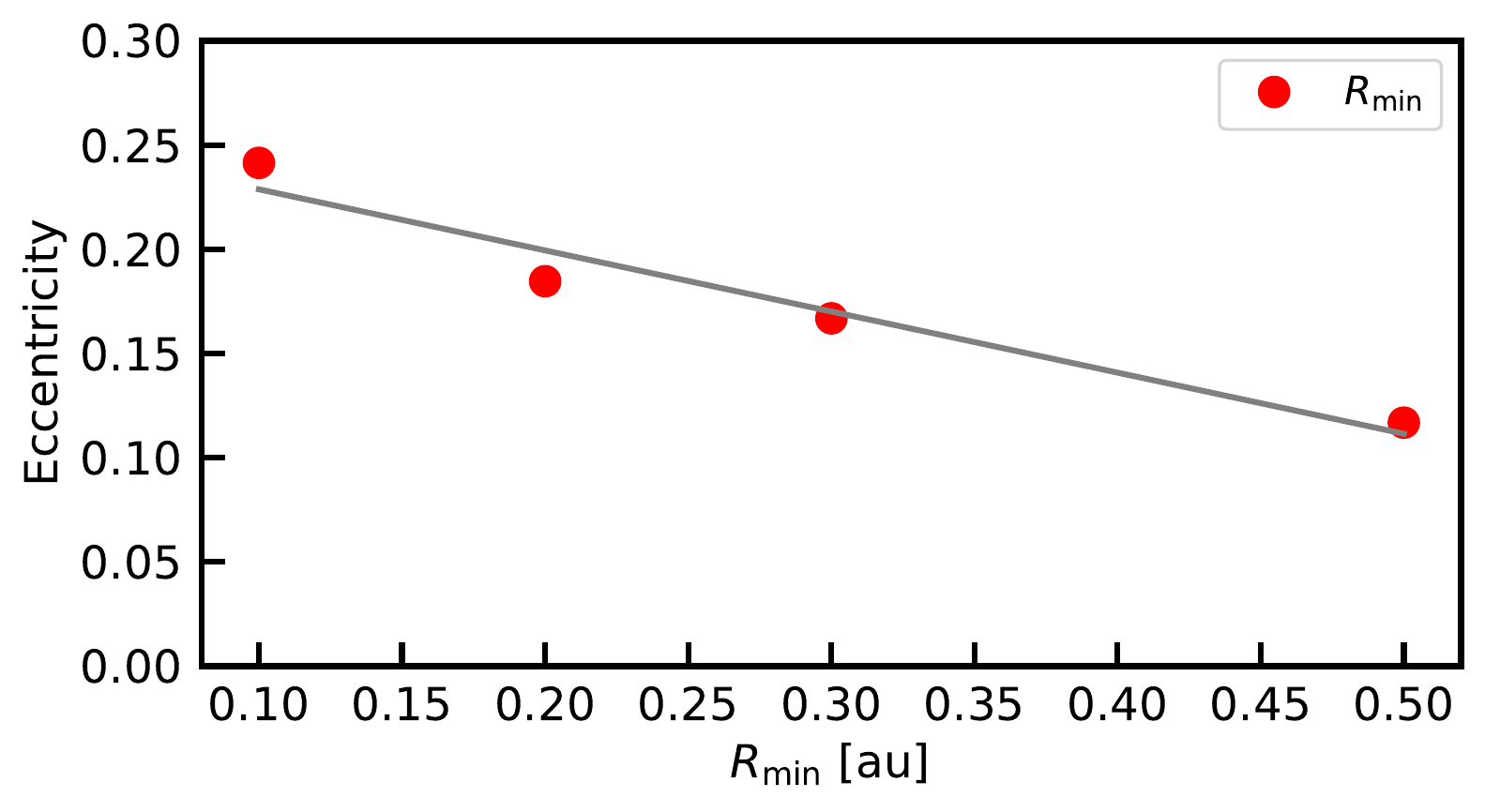}
	\caption{Time-averaged mass-weighted disk eccentricity plotted against the
	location of the inner domain boundary.}
	\label{fig:log_inner_radii_ecc}
\end{figure}
\begin{figure}[!t]
	\centering
	\includegraphics[width=88mm]{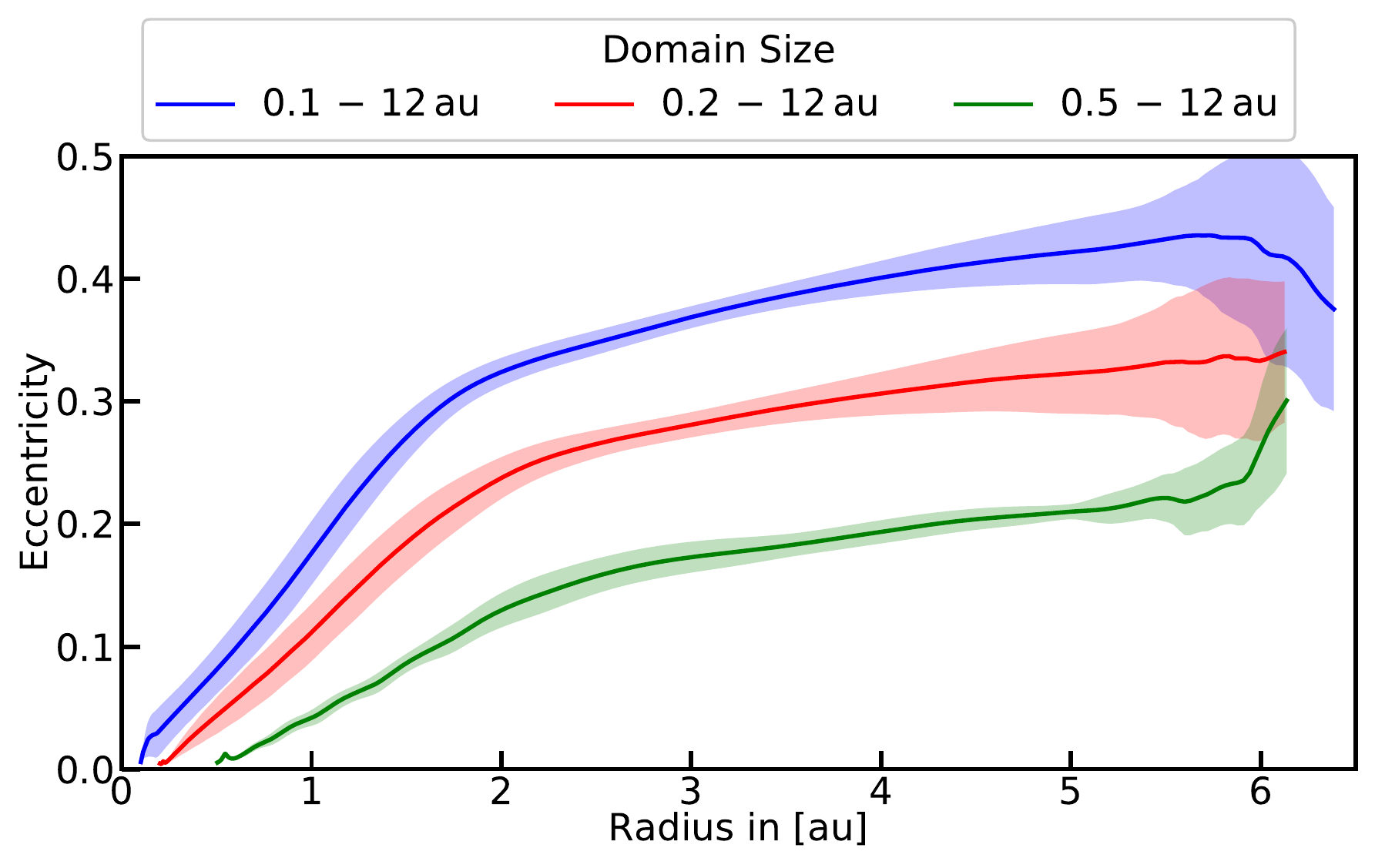}
	\caption{Azimuthally averaged disk eccentricity  (see
	Eq.\,\eqref{eq:phi-average})  for different locations of the inner radius.
	Solid lines are averaged over 200 snapshots taken at the binary apastron
	during the simulation time from $1000\,T_\mathrm{bin}$ to
	$1200\,T_\mathrm{bin}$. The shaded areas show the $1\sigma$ variations. The
	disk has a radius of $\approx5\,\mathrm{au}$, eccentricities at radii beyond
	that do not contribute to the disks eccentricity.}
	\label{fig:time_avg_log_inner_radii}
\end{figure}
\begin{figure*}[!htb]
	\centering
	\includegraphics[width=180mm]{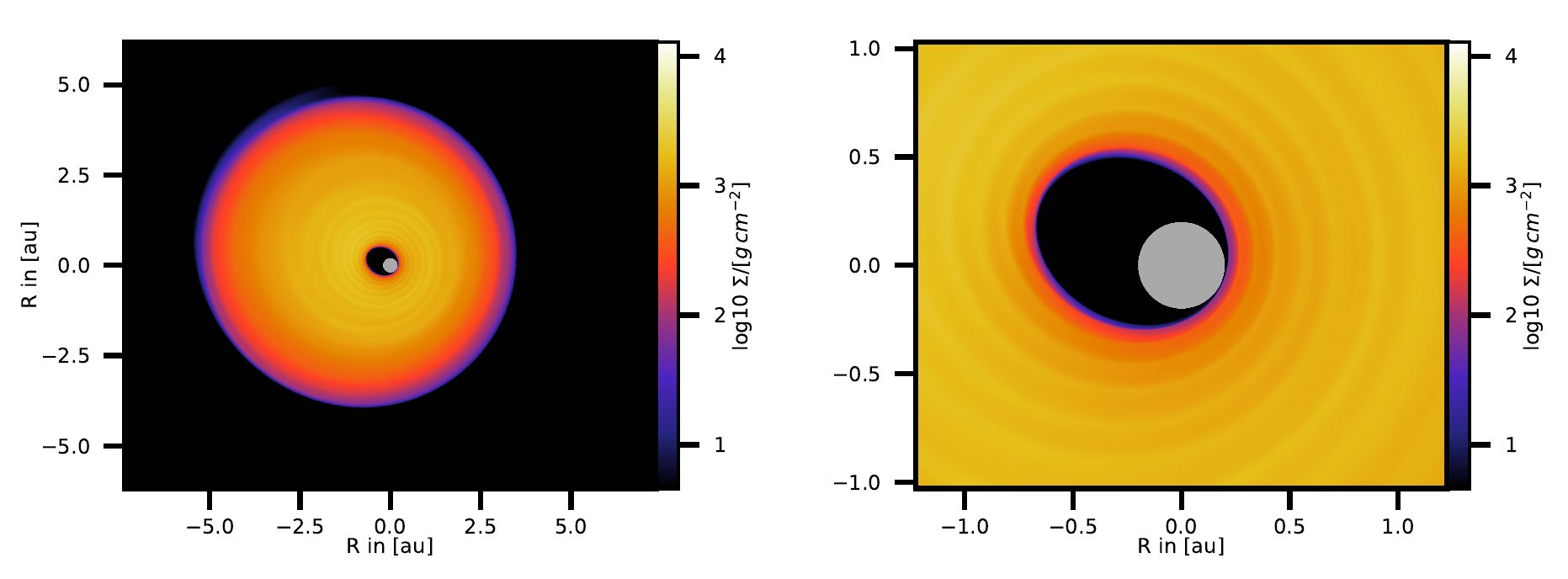}
	\caption{Snapshot of the disks surface density after 481 orbital periods at
	the binary apastron. The simulation used an open inner boundary condition.
	Right: zoom in on the inner $1\,\textrm{au}$. Mass being removed at the
	inner boundary (gray area) causes the development of an eccentric hole.}
	\label{fig:open_boundary_dens}
\end{figure*}

\subsection{Location of the outer boundary}
Previous studies on $\gamma$-Cephei used domain-sizes of $0.5\,\mathrm{au} -
8\,\mathrm{au}$, equivalent to .025\,$a_\mathrm{bin} - 0.4\,a_\mathrm{bin}$,
\citep{kley2008planet,muller2012circumstellar,paardekooper2008planetesimal},
while \cite{marzari2009eccentricity} used $0.5\,\mathrm{au} - 15\,\mathrm{au}$
but with a larger $a_\mathrm{bin} = 30\,\mathrm{au}$, instead of our
$a_\mathrm{bin} = 20$. To check the impact of the chosen value of the outer
boundary on the disk dynamics, we ran simulations using different values for
$R_\mathrm{max}$. We found that the outer radius $R_\mathrm{max}$ has little
effect on the dynamics of the disk, but does change the mass loss rate across
the outer boundary.
When the companion passes periastron, mass is ejected from the primary disk. If
$R_\mathrm{max}$ is too small, mass ejected from the disk is removed from the
simulation before it can be re-accreted onto the primary disk. Increasing
$R_\mathrm{max}$ further than $12\,\mathrm{au}$ (binary distance at periastron)
leads to the same mass loss rates once the disk has reached equilibrium. Hence,
the simulation domain should therefore at least contain the whole Roche lobe of
the primary in apocenter, as in our standard case.
\subsection{Location of the inner boundary}
The location of the inner domain radius, $R_\mathrm{min}$, directly influences
the dynamics of the disks. The further out the position of the inner boundary,
the lower the disk eccentricity with an approximately linear dependency, see
\figref{fig:log_inner_radii_ecc}. For reflective boundaries, the radial velocity
and hence eccentricity is forced to zero at the boundary. If the boundary is
closer to the star, it has less influence on the rest of the disk, allowing it
to develop higher eccentricity, see \figref{fig:time_avg_log_inner_radii}. This
increase in eccentricity for smaller $R_\mathrm{min}$ could also be explained by
additional higher-order resonances still being captured inside the domain
\citep{marzari2009eccentricity}. Similar to the disk eccentricity, the
precession rate increases for decreasing $R_{min}$, from
$0.09\,T_\mathrm{bin}^{-1}$ at $R_{min} = 0.5\,\mathrm{au}$ to
$0.12\,T_\mathrm{bin}^{-1}$ at $R_{min} = 0.1\,\mathrm{au}$. Therefore, adding
an inner disk region influences the eccentricity and precession rate of the
whole disk while the radial temperature and the surface density profile remain
unchanged (image not shown).

For open boundaries, a smaller $R_\mathrm{min}$ reduces the artificial mass loss
at the inner boundary. Due to the high computational cost of reducing the inner
radius of the domain, we limit $R_\mathrm{min}$ to $0.2\,\mathrm{au}$ for the
rest of our simulations.
\begin{figure}[!htb]
	\centering
	\includegraphics[width=88mm]{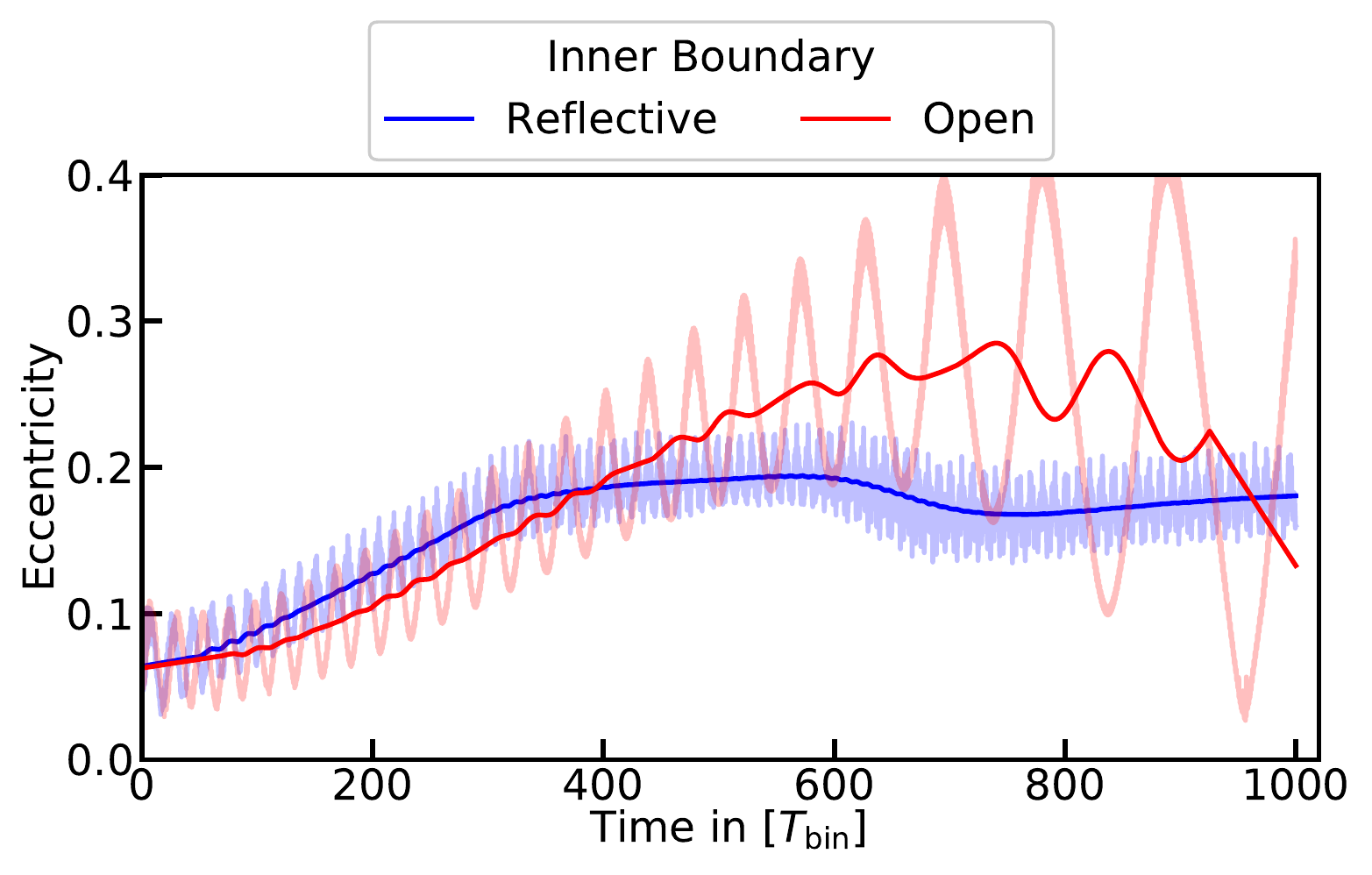}
	\caption{Time evolution of the mass-weighted disk eccentricity for different
    inner boundary conditions. Solid lines are averaged values while the
    transparent lines are the simulation data. The simulation with an open
    boundary at the inner domain does not reach an equilibrium state due to the
    rapid mass loss. The lower temperatures from losing mass slow down the
    precession rate.}
	\label{fig:log_inner_boundary_ecc}
\end{figure}
\subsection{Inner boundary condition}
\begin{figure}[!htb]
	\centering
	\includegraphics[width=88mm]{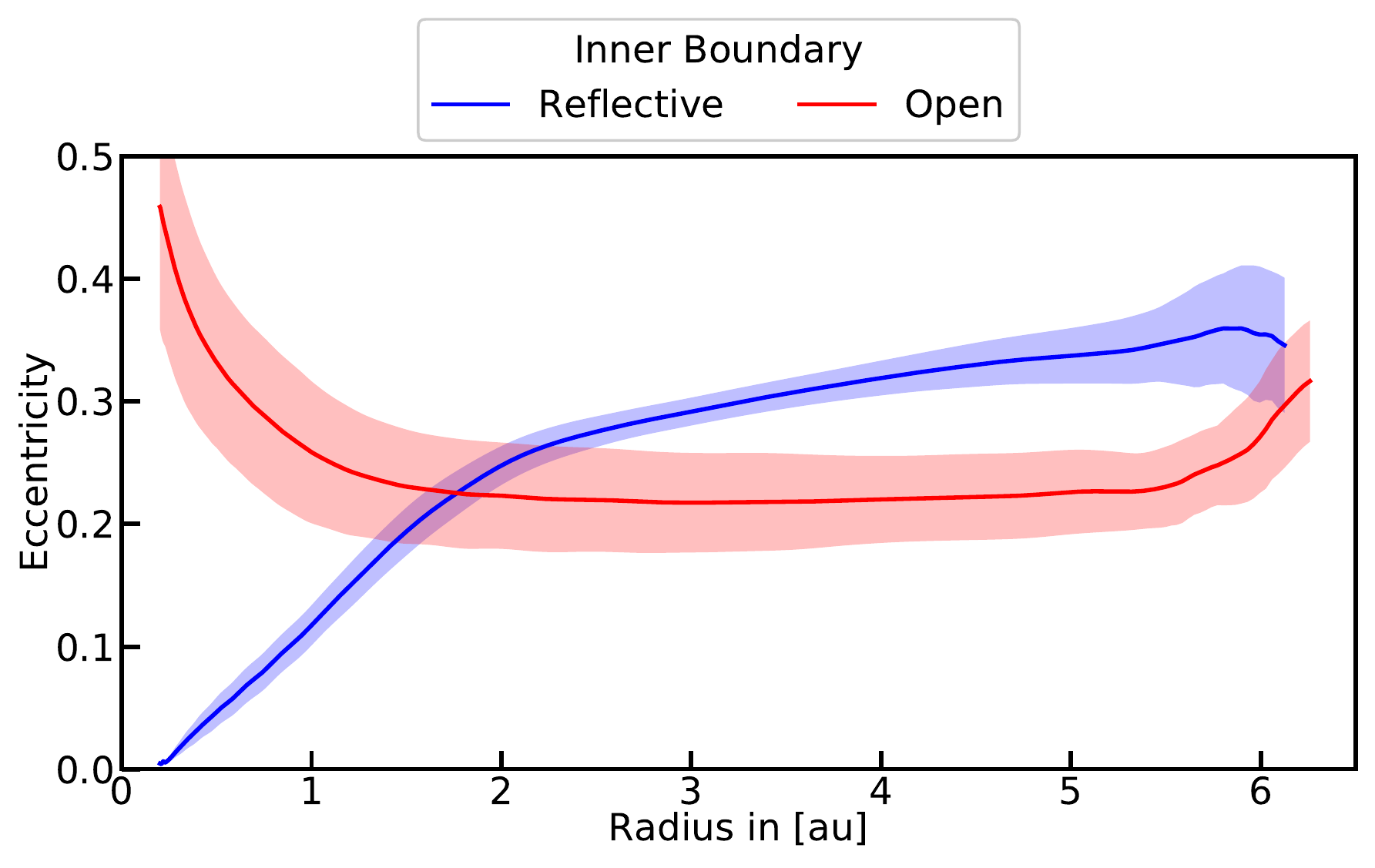}
	\caption{Radial profiles of the time averaged mass-weighted disk
    eccentricity for different inner boundary conditions. Solid lines are
    averaged over 200 snapshots taken at binary apastron between $T = 400$ and
    $600T_\mathrm{bin}$. The shaded areas show the $1\,\sigma$ variations.
    Eccentricity as well as precession are more uniform throughout the disk when
    open boundaries are used. The radial profiles are cut at a radius where the
    time averaged surface density drops below 100 times the floor value.}
    \label{fig:time_avg_log_inner_boundary_ecc}
\end{figure}

We also ran a test using an open inner boundary condition without damping at the
inner boundary. The simulation develops a similar eccentricity compared to the
standard case with a reflective inner boundary condition
(\figref{fig:log_inner_boundary_ecc}). Since any mass that crosses the inner
boundary is removed from the simulation, the disk loses mass very fast and forms
an elliptic inner hole (\figref{fig:open_boundary_dens}). This behavior is in
agreement with the simulations shown in \cite{kley2008simulations} and
\cite{marzari2012eccentricity}. This rapid mass loss prevents the disk from
reaching an equilibrium state as the disk continuously becomes colder, which
causes the precession rate to slow down (compare \figref{fig:precession_tmep}
below). The disk's eccentricity profile behaves differently. Throughout the
whole disk, the eccentricity is constant, whereas in the simulation with
reflective boundaries, the eccentricity is forced to zero at the inner boundary
and grows outward (\figref{fig:time_avg_log_inner_boundary_ecc}). Similarly, the
disk precesses as a solid body at all times, while for simulation with
reflective boundary, we find deviations from solid body precession, see
\secref{sec:ecc_prec} for more details. The gas at the edge of the inner
eccentric hole develops high radial velocities and variations in the azimuthal
velocity that severely limit the timestep, making these simulations
computationally too expensive for a full parameter study in this work.

\section{Dependence on physical parameters}
\label{sec:physical}
Having analyzed the impact of purely numerical parameters, we conclude, that the
computational setup of the standard model (see Table\,\ref{table:parameters})
seems to be sufficient to capture all relevant physical properties of the disks.
We shall use it throughout this section to study the effects of physical
parameters on the disk's dynamics. The $\alpha$-parameters and initial disk
masses tested in this section are listed above in
\tabref{table:parameter_range}.
\begin{table}[!t]
	\centering
	\caption{List of all the simulations for the parameter study.
	\label{table:parameter_range}}
\renewcommand{\arraystretch}{1.2}
\begin{tabular}{l l l l l} 
	\hline\hline 
	$\alpha$-parameter & $1\cdot10^{-4}$ & $3\cdot10^{-4}$ & \color{blue}$\underline{1\cdot10^{-3}}$ &
	$3\cdot10^{-3}$ \\
	& $5\cdot10^{-3}$ & $6\cdot10^{-3}$ & $7\cdot10^{-3}$ & $1\cdot10^{-2}$ \\
	\hline 
	Initial disk mass & $2\cdot10^{-3}$ & $5\cdot10^{-3}$ & \color{blue}$\underline{1\cdot10^{-2}}$ &
	$2\cdot10^{-2}$\\
	in $[\mathrm{M}_{\sun}]$ & $3\cdot10^{-2}$ & $4\cdot10^{-2}$ &
	$5\cdot10^{-2}$ &\\
	\hline
\end{tabular}
	\tablefoot{The simulations use the same parameters as the fiducial model in
		\tabref{table:parameters}, but have one disk parameter changed. Each
		number represents one simulation. The default parameters are underlined
		in blue.}
\end{table}
\subsection{Disk mass}
The disk's mass density has a direct influence on its temperature, first by
increasing the heating through the viscous stress tensor and secondly by
increasing the optical depth thereby reducing the radiative cooling efficiency.
Thus, increasing the disk mass should directly increase the disk temperature and
subsequently impact the disk dynamics, that is, its eccentricity. We tested this
by varying the initial disk masses, ranging from $2\cdot10^{-3}$ to
$5\cdot10^{-2}\,M_\odot$, while keeping the other parameters unchanged from the
standard model. 

\begin{figure}[!t]
	\centering
	\includegraphics[width=88mm]{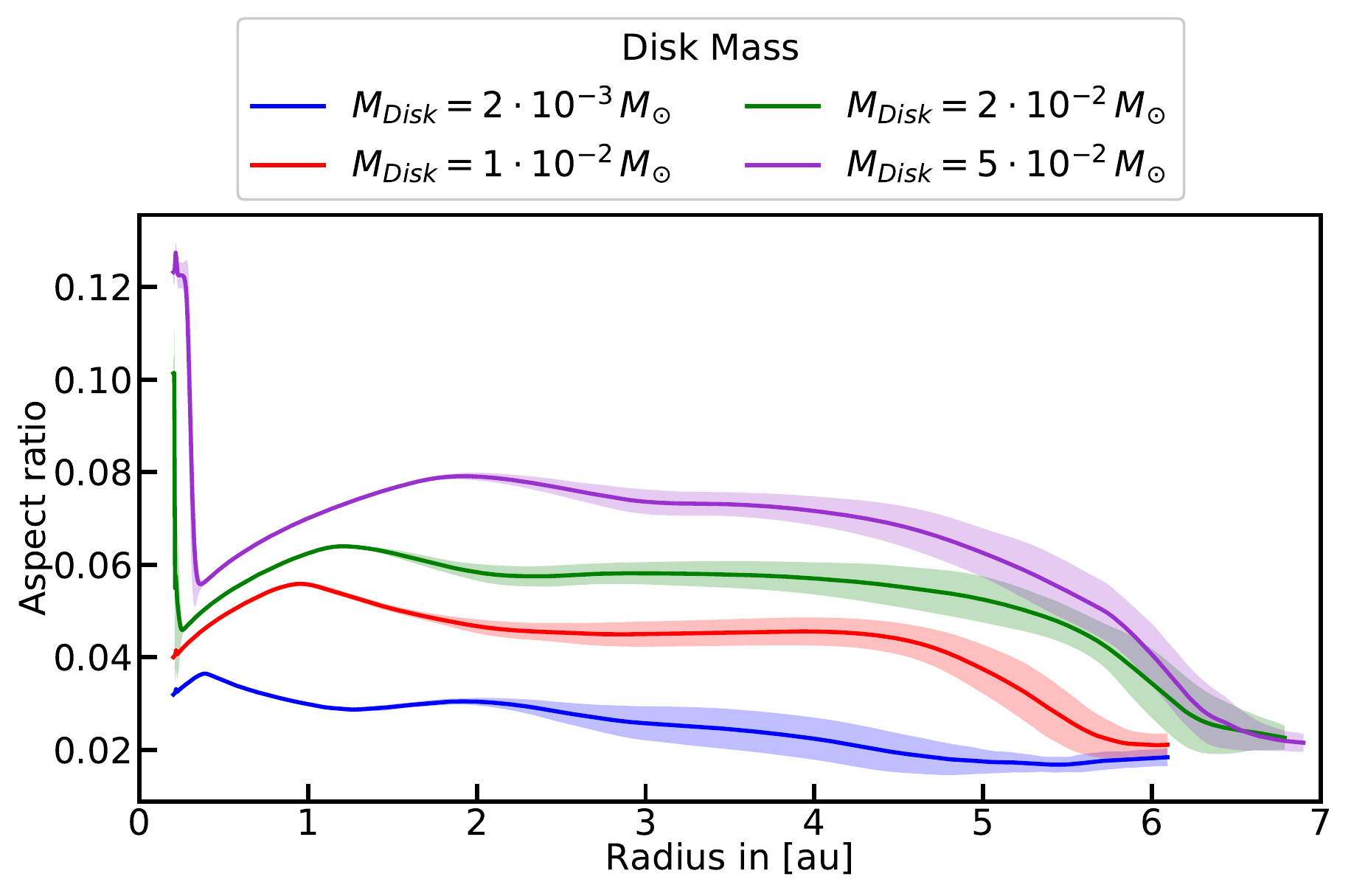}
	\caption{Radial profiles of the time averaged mass-weighted disk scale
	height, here $h=H/r$, for different initial disk masses. Solid lines are
	averaged over 200 snapshots taken at binary apastron during the simulation
	time from $1300\,T_\mathrm{bin}$ to $1500\,T_\mathrm{bin}$. The shaded areas
	show the $1\,\sigma$ variations. The radial profiles are cut at a radius
	where the time averaged surface density drops below 100 times the floor
	value.}
	\label{fig:average_mass_h}
\end{figure}

\subsubsection{Temperature profile}
The radial disk temperature (here given as the relative disk thickness)
resulting from different disk masses is displayed in
\figref{fig:average_mass_h}, and it clearly shows a monotonic increase in aspect
ratio with disk mass within the disk proper, that is, inside of about
$6\,\mathrm{au}$. But the individual curves show some substructures. For the
lowest mass ($2 \cdot 10^{-3} M_\odot$) case (the blue line in
\figref{fig:average_mass_h}), a drop-off in aspect ratio is visible for radii
inside of $\sim 2\,\mathrm{au}$. Inside of this radius, the disk temperature
surpasses the ice sublimation temperature $\sim 160\,\mathrm{K}$. This trend
reverses when all the ices are sublimated at $\sim220\,\mathrm{K}$ (at around
$1.2\mathrm{au}$) and the opacity increases with temperature again. 

Not considering, for now, the strong increase in $h$ close to the inner
boundary, all models show a drop in aspect ratio toward the inner edge, which
indicates the sublimation of dust grains at about $1100\,\mathrm{K}$ and a
lowering of the opacity. With increasing mass, this transition moves further
out, as does the maximum of $h$. The standard model with $M_\mathrm{d} =
10^{-2}M_\odot$ reaches the threshold at $ \sim 1\,\mathrm{au}$ while for the
highest disk mass it lies at about $2\,\mathrm{au}$. At the very inner edge, the
models with larger disk mass show a strong increase in temperature up to over
$3300\,\mathrm{K}$. This is caused by the dissociation of hydrogen, which
drastically increases the opacity, leading to high aspect ratios.

We also tested the effect stellar irradiation from the central star would have
on our disks using the irradiation model from \citet{menou2004lowmass}.
Following the procedure described in \citet{ziampras2020wakes}, we took a
time-averaged radial density profile of our disks and calculated its temperature
in thermal equilibrium under the influence of viscous heating, stellar
irradiation, and radiative cooling. We found that stellar irradiation does not
affect on the optically thick, dense inner half of the disk  ($r \lesssim 2
\text{--}\,4\,\mathrm{au}$ depending on model) and only becomes relevant at the
outer parts. However, due to the existence of a maximum in the aspect ratio in
this region (see \figref{fig:average_mass_h}), the outer parts would be shadowed
by the inner regions (from the star or hot inner disk) if a more realistic
irradiation model was used. We conclude that stellar irradiation has no effect
on the dynamics of our disks and can be neglected.
\subsubsection{Time evolution of the cccentricity}
\begin{figure}[!htb]
	\centering
	\includegraphics[width=88mm]{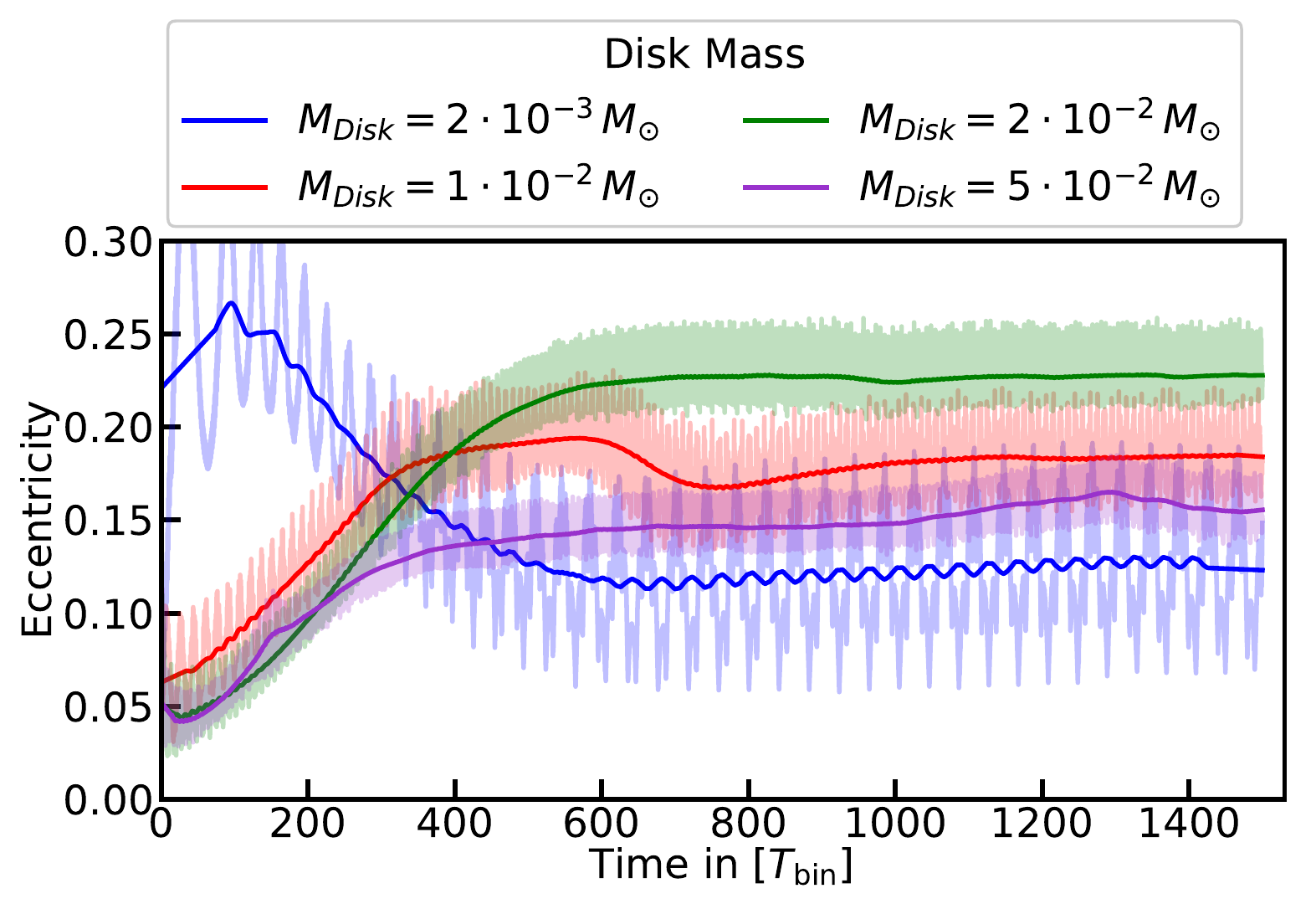}
	\caption{Time evolution of the mass-weighted disk eccentricity for different
	initial disk masses.}
	\label{fig:log_mass_ecc}
\end{figure}

\begin{figure}[!htb]
	\centering
	\includegraphics[width=88mm]{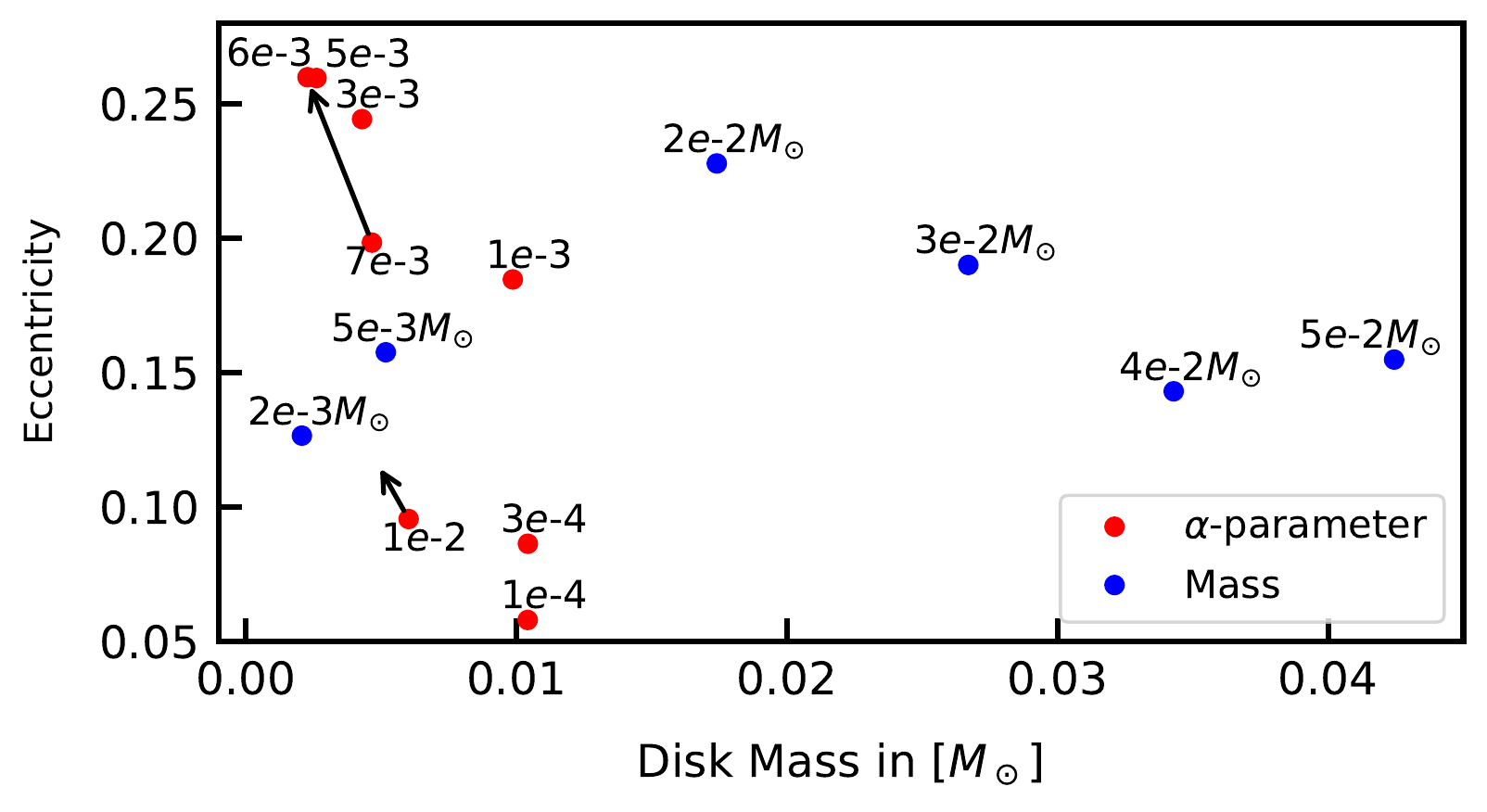}
	\caption{Time-averaged mass-weighted disk eccentricity plotted against the
	mass-weighted disk temperature. The arrows indicate the state after
	$2000\,T_\mathrm{bin}$. At that time the $\alpha = 10^{-2}$ case has not
	reached an equilibrium yet.}
	\label{fig:ecc_mass}
\end{figure}

The time evolution of the global disk eccentricity is displayed in
\figref{fig:log_mass_ecc} for four different disk masses. After about 600 to 800
$T_\mathrm{bin}$ all models settle to a quasi-stationary state. The simulations
show that the small eccentricity for the low mass case is accompanied by larger
deviations from solid body precession. Deviations from solid body precession
appear as varying precession rates at different radii inside the disk. In our
simulations, we observe that the disk's eccentricity is suppressed, when
neighboring gas rings precess independently of each other. From
\figref{fig:log_mass_ecc} it can also be seen, that the oscillation period of
the eccentricity becomes shorter with higher disk mass, which implies a faster
retrograde precession rate. We discuss these effects in more detail in
\secref{sec:ecc_prec}.
\linebreak \indent The low mass case $2\cdot10^{-2}\,M_\odot$ (blue line in
\figref{fig:log_mass_ecc}) has a peculiar eccentricity evolution because it is
the one farthest away from an equilibrium profile during initialization. It is
too cold to precess for the first 20 binary orbits; because the longitude of
pericenter is positive in the beginning, this results in a rapid eccentricity
growth (compare the eccentricity growth for $\omega_\mathrm{d} > 0$ in
\figref{fig:peri_ecc}). The disk becomes more compacted by the tidal forces of
the companion and starts precessing, resulting in a hot inner region and cold
outer rim with a large temperature gradient across the disk. Due to the strong
temperature gradient, the disk precesses at different speeds at different radii,
resulting in strong damping of the eccentricity, as mentioned above.
\linebreak \indent The resulting disk eccentricities in the equilibrium state
are shown in \figref{fig:ecc_mass}, which includes additionally the results for
different viscosities. We find that the simulation with a disk mass of
$2\cdot10^{-2}\,M_\odot$ (resulting in an aspect ratio of $h = 0.057$) develops
the largest eccentricity. More massive and less massive disks develop a smaller
eccentricity (see blue points in \figref{fig:ecc_mass}). This finding is in
agreement with \cite{marzari2012eccentricity} who found their eccentricities
peaking at $h = 0.05$ in their locally isothermal runs (compare their Fig.\,1).
Their explanation states that for cold disks, the perturbations (spiral arms)
move inward more slowly and get tightly wound up. They are then more affected by
viscous damping and become less efficient at depositing angular momentum in the
inner parts of the disk. In hot disks, radiative damping dissipates the
perturbations faster, leading to an overall lower disk eccentricity. The
drop-off in eccentricity when deviating from the optimal $h$ with the highest
$e_\mathrm{d}$ is much smaller in our simulations, with the time-averaged
eccentricity never dropping below $e_\mathrm{d} \sim 0.12$.
\linebreak
\subsubsection{Mass loss}
As explained above, the disks lose mass across the outer boundary of the
computational domain. Starting from the initial setup, all models typically lose
about $2.5\%$ of their mass during the first few orbits. Ignoring this mass loss
at the beginning, the disks lose mass on a longer timescale which is mostly
ejected during the binary periastron passage when the disk's longitude of
pericenter is aligned to the binary apastron, see Fig.\,\ref{fig:timeline}
above. Mass outflow only occurs if the disks are large and eccentric enough to
overflow the primary's Roche lobe during the periapse. It regulates itself
because at some point, the disks become small enough to fit entirely inside the
primary's Roche lobe and mass loss during pericenter passage becomes negligible.
The mass-loss rate increases during the growth of the disk's eccentricity and
then settles to a stationary value. For the standard case, this
quasi-equilibrium is reached at around 600$T_\mathrm{bin}$. After that, the disk
loses mass at a rate of about $3.6 \cdot 10^{-9} M_\odot/$yr for the standard
case, and $1.4\cdot10^{-7} M_\odot/$yr for the more viscous
$\alpha=6\cdot10^{-3}$ case. Disks with a viscosity smaller than our standard
case do lose even less mass to outflow, while our most viscous disks lose 80\%
of their mass within our simulation time of 1500 binary orbits, see
\figref{fig:log_viscosity_mass} below.

In the simulation with an open inner boundary, the disk loses 80\% of the
initial mass within 900 binary orbits. Due to its low viscosity of $\alpha =
10^{-3}$, very little mass is lost by outflow, and most of the mass is lost
through the inner boundary. This mass loss might be enhanced by our artificial
large inner boundary. The mass loss through the inner boundary can be described
as an exponential decay, as would be expected from the viscous accretion around
a single star, and starts at $2.1 \cdot 10^{-7} M_\odot/$yr when the disk still
has its full mass and drops to $6.5 \cdot 10^{-8} M_\odot/$yr at
$900\,T_\mathrm{bin}$ when the disk mass is at $2.5\cdot10^{-3}\,M_\odot$.

\subsection{Viscosity}
\begin{figure}[!htb]
	\centering
	\includegraphics[width=88mm]{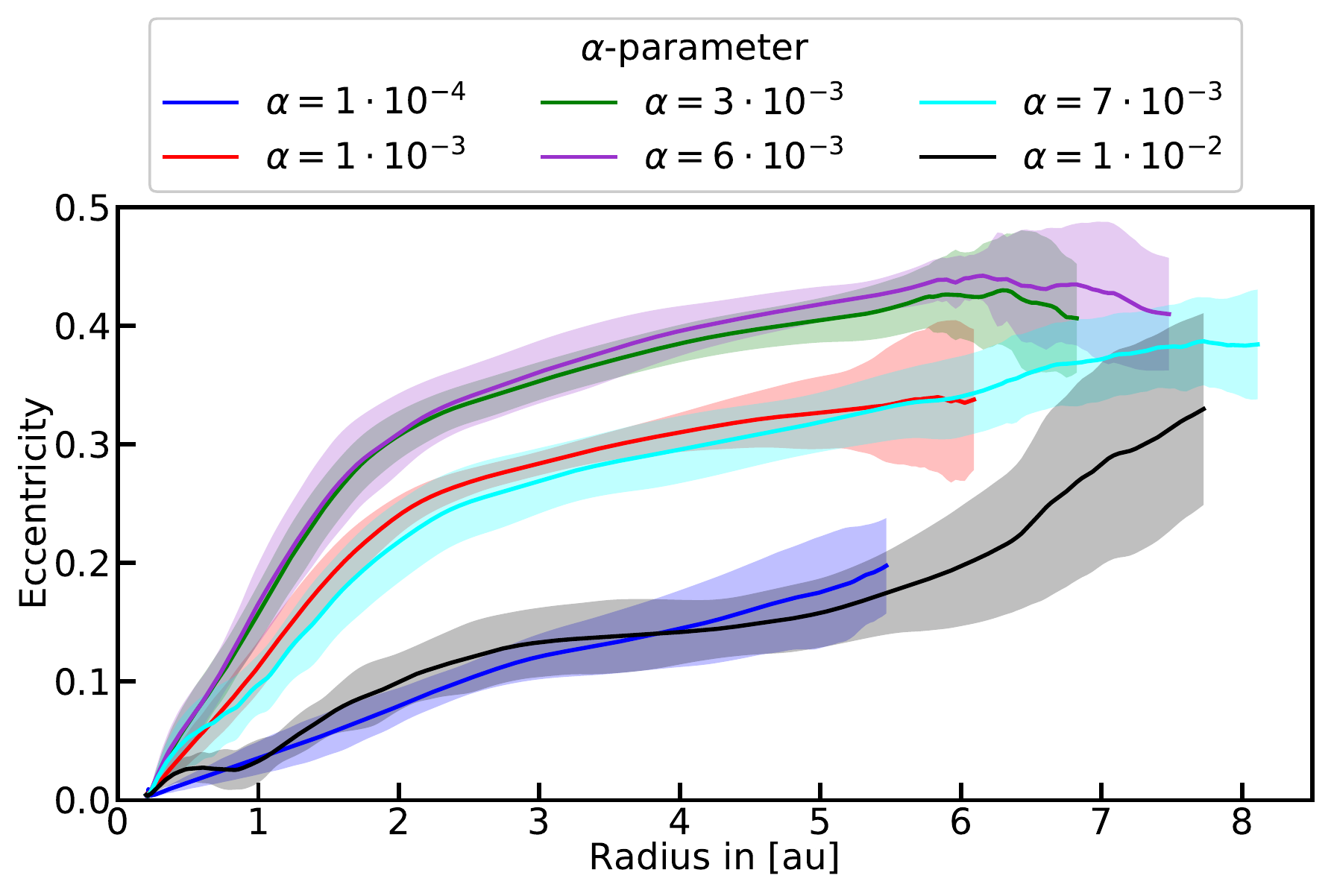}
	\caption{Radial profiles of the time averaged mass-weighted disk
	eccentricity for different $\alpha$ values. Solid lines are averaged over
	200 snapshots taken at the binary apastron during the simulation time from
	$1300\,T_\mathrm{bin}$ to $1500\,T_\mathrm{bin}$. The shaded areas show the
	$1\,\sigma$ variations. The radial profiles are cut at a radius where the
	time averaged surface density drops below 100 times the floor value.}
	\label{fig:time_avg_log_viscosity_ecc}
\end{figure}

We tested $\alpha$ values ranging from $10^{-4}$ to $10^{-2}$. The resulting
eccentricities as a function of radius are displayed in
\figref{fig:time_avg_log_viscosity_ecc}. For smaller viscosities, from $\alpha =
10^{-4}$ to $6 \cdot 10^{-3}$, we confirm the trend found in
\cite{kley2008simulations} and \cite{muller2012circumstellar} that higher
viscosity leads to more eccentric disks. 

For even higher $\alpha$, the disk dynamics evolve differently, and eccentricity
drops again, as indicated by the cyan and black curves in
\figref{fig:time_avg_log_viscosity_ecc}. For these two highest $\alpha$-values
($\alpha = 7\cdot10^{-3}$ and $10^{-2}$) the disks first enter a quiet state
that is steady, low-eccentric and non-precessing. Only after several hundreds of
orbits do the disks start to precess and become more eccentric. At the time of
$2000\,T_\mathrm{bin}$ the $\alpha = 7\cdot10^{-3}$ simulation reaches the same
final eccentricity of $\bar{e}_d \approx 0.25$  as the $\alpha = 6\cdot10^{-3}$
case, while the simulation with $\alpha = 10^{-2}$ continues to grow its
eccentricity at a slow but constant rate.

\begin{figure}[!htb]
	\centering
	\includegraphics[width=88mm]{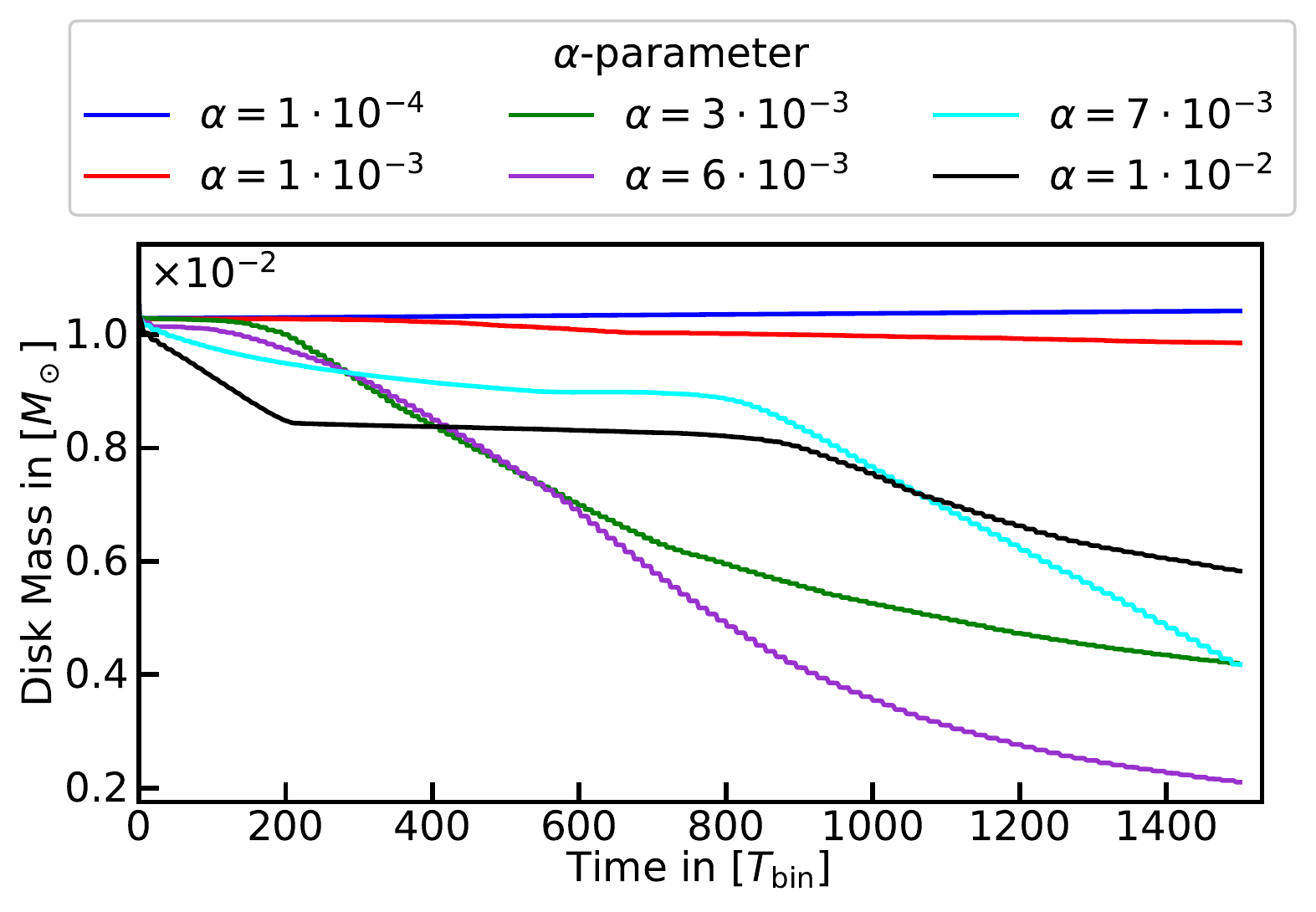}
	\caption{Time evolution of the disk mass for different $\alpha$ values.}
	\label{fig:log_viscosity_mass}
\end{figure}

Higher $\alpha$ values result in stronger viscous torques, causing the disk to
spread further outward. A larger disk is more affected by the companion and thus
develops a larger eccentricity. The gas at the outer rim of the disk also has
the largest eccentricity. Because of that, it is ejected from the disk during
periastron passage which prevents further growth of eccentricity and limits the
disk radius. The evolution of disk mass for the different disk viscosities is
shown in \figref{fig:log_viscosity_mass}, which shows a clear trend for the
lower viscosity cases that the mass loss increases with viscosity, as expected.
At first, the two high viscosity cases show relatively small mass loss against
the initial expectation. This is due to the disks initially remaining much more
circular and only start losing mass after becoming excited, in the same way as
described above.
\section{Global trends}
\label{sec:ecc_prec}
In this section, we look for global trends in the simulations from the previous
section. We vary simultaneously, mass and viscosity of the disks and we present
the data in different forms to better visualize the connection between the
physical state of the disk and its dynamical behavior.
\subsection{Eccentricity}
\begin{figure}
	\centering
	\includegraphics[width=88mm]{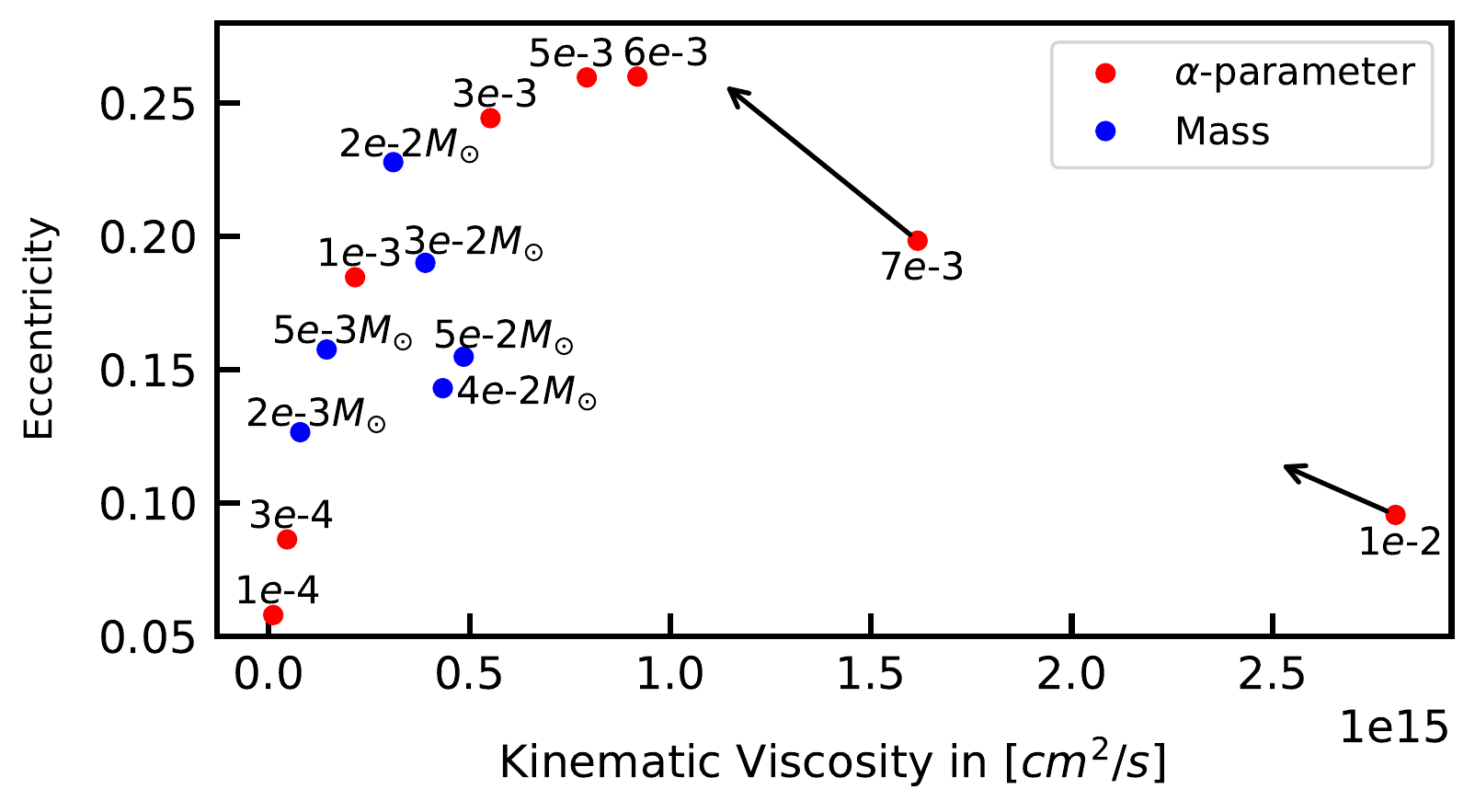}
	\caption{Time-averaged mass-weighted eccentricity plotted against the time
	averaged mass-weighted viscosity. The arrows indicate the state after
	$2000\,T_\mathrm{bin}$. At that time the $\alpha = 10^{-2}$ case has not
	reached an equilibrium yet.}
	\label{fig:ecc_vis}
\end{figure}

\begin{figure}[!htb]
	\centering
	\includegraphics[width=88mm]{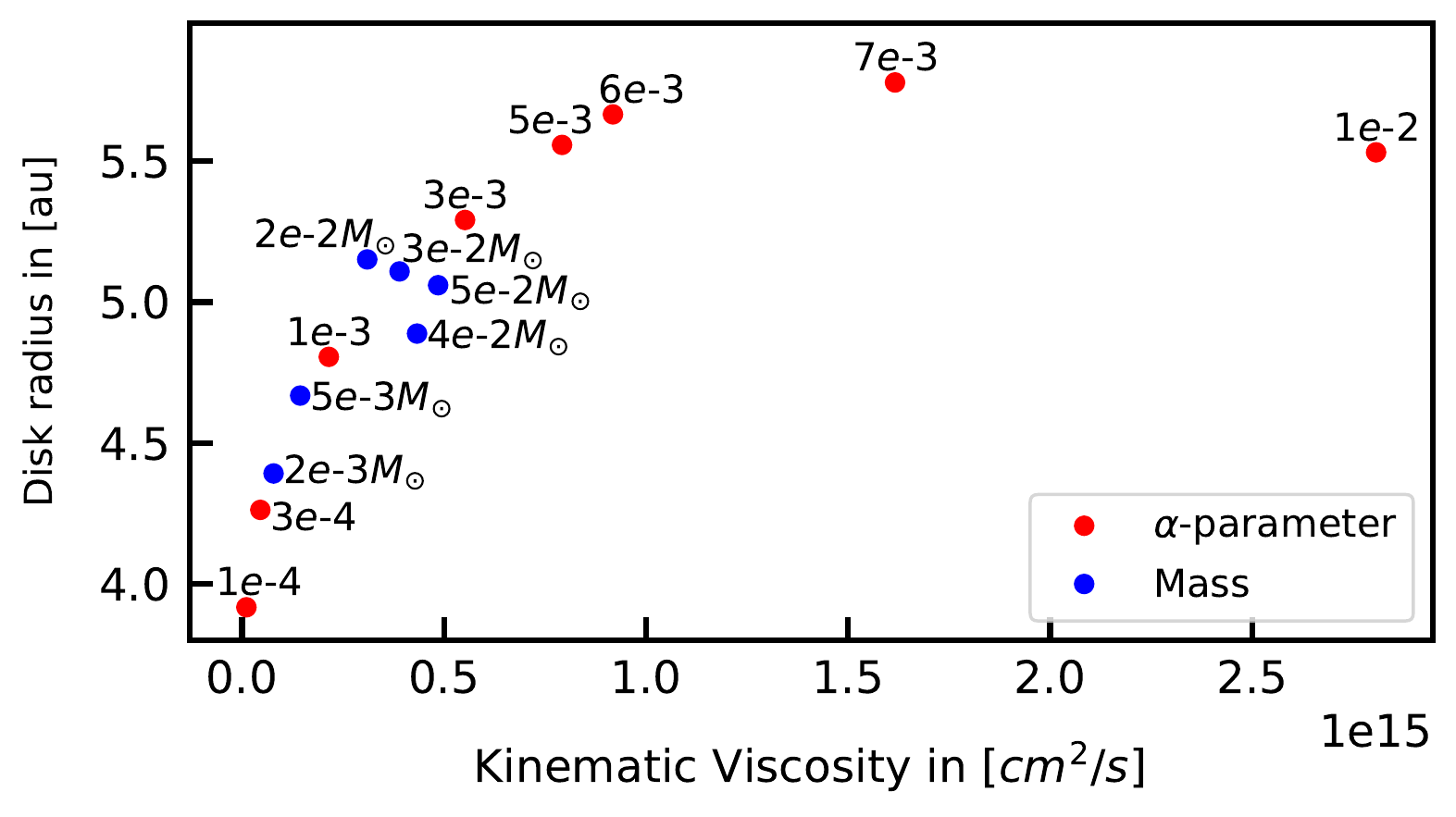}
	\caption{Time-averaged disk radius plotted against mass-weighted kinematic
	viscosity.}
	\label{fig:vis_radius}
\end{figure}

Figures\,\ref{fig:ecc_vis} and \ref{fig:vis_radius} show the averaged disk
eccentricity and disk radius plotted against the global averaged disk viscosity.
The data is first mass-averaged, see Eq.\,\eqref{eq:global-average}, and then
time-averaged over the final ten precession periods, at simulation times between
1200 and $1500\,T_\mathrm{bin}$. We measure the precession periods as the time
between two adjacent peaks of the longitude of pericenter (see
\figref{fig:peri_ecc}). The blue dots always represent data from simulations
with different disk masses and red dots data from those with different
viscosity. Figure\,\ref{fig:ecc_vis} shows a positive correlation between
viscosity and eccentricity, except for the high viscosity simulations. These
start in a quiet state and retain a smaller eccentricity for the whole run time.
In the following discussion, we ignore those two simulations because they have
not reached an equilibrium state at the end of our standard simulation time of
$1500\,T_\mathrm{bin}$. The positive correlation between eccentricity and
viscosity is in agreement with earlier studies
\citep{kley2008simulations,muller2012circumstellar}.  It can be explained by
larger viscous torques increasing the disk's size, making it more susceptible to
perturbations of the companion and causes a higher eccentricity.
The simulations with different disk masses follow the same trend, except for the
higher mass simulations, see \figref{fig:ecc_mass} above. We attribute the lower
eccentricity in high mass simulation to radiative damping of the perturbations.

Similarly, diffusive action of viscosity is reflected in \figref{fig:vis_radius}
which shows a positive correlation between disk size and viscosity. The disk
expansion due to viscosity stops at an average disk radius of $\sim
5.7\,\mathrm{au}$, equivalent to 0.28 $a_\mathrm{bin}$, see
\figref{fig:vis_radius}, and a mass-weighted eccentricity of $0.25$,
\figref{fig:ecc_vis}. It is important to note that since we keep the number of
grid cells constant, simulations with a low aspect ratio are less resolved. This
could enhance the drop-off in eccentricity for simulations with low viscosity or
low mass. 

We find that the disk's eccentricity increases linearly with its radius, which
has been noted by \citet{marzari2009eccentricity}, who changed the disk size by
varying the binary's eccentricity. We note that the linear correlation between
size and eccentricity could also partly be explained by our method of
determining the disk's radius as the distance from the primary that contains
99\% of the disk mass.

\subsection{Precession rate}
\begin{figure}[!htb]
	\centering
	\includegraphics[width=88mm]{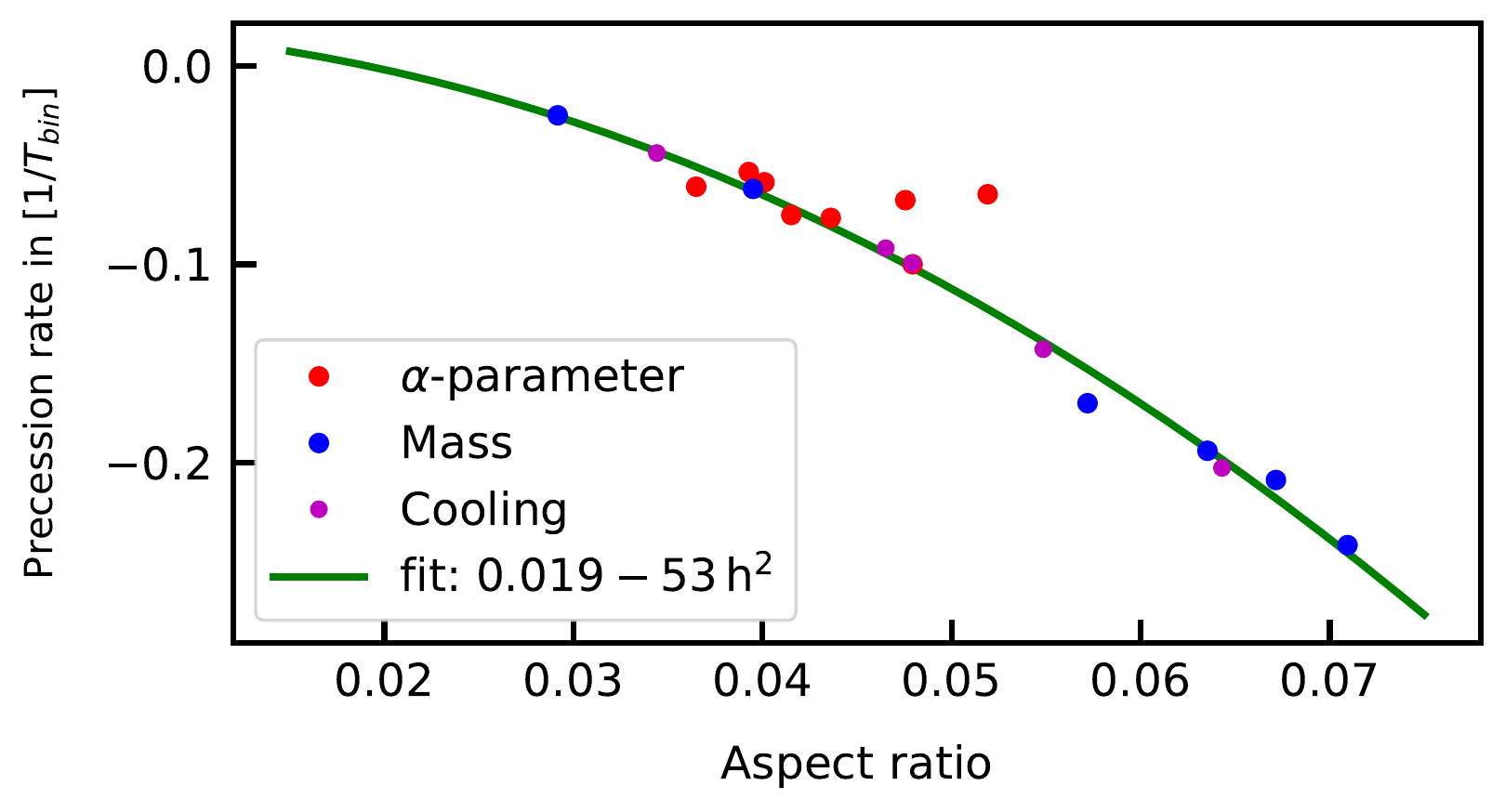}
	\caption{Precession rate plotted against the mass-weighted disk aspect
    ratio. Red dots mark the simulations with different $\alpha$ values, blue
    dots with different disk masses and magenta dots with different cooling
    factors and lower resolution. The green line is a quadratic fit through the
    data points. The precession rate depends linearly on the disk Temperature
    with $T \propto h^2$.}
	\label{fig:precession_tmep}
\end{figure}

The precession rate was determined as described in the previous section. We find
a slow retrograde precession for all disks, with one precession period lasting
between $5$ to $40$ binary orbital periods. The precession rates are shown in
\figref{fig:precession_tmep} as a function of the global disk aspect ratio. The
models for different disk masses and viscosities show a unique quadratic
relation, fitted approximately by
\begin{equation}
  \label{eqn:disk_precession}
    \dot{\omega}_\mathrm{d} = ( 0.019 - 53\,h^2 ) T_\mathrm{bin}^{-1} \,.
\end{equation}
In general, the disk's precession rate depends on the balance of gravitational
and pressure forces \citep{goodchild2006dynamics}. The gravitational forces
induce a prograde precession while the pressure forces result in retrograde
precession. We find retrograde precession for all our simulations, indicating
that the disks are hot enough to be pressure-dominated. Our values for the
precession rate and, in particular, the quadratic dependence with the mean scale
height $h$ are in excellent agreement with the locally isothermal simulations in
\cite{kley2008simulations}, obtained for circular binaries. 

Since in our model $T \propto h^2$, this implies a linear relation between disk
temperature and precession rate. We restarted our standard case with a modified
heating and cooling rate to test the linear relation between precession rate and
temperature. The heating and cooling rate were directly multiplied by factors of
either 0.1 or 10. After that, the precession rates adjusted themselves on the
thermal timescale, which is only a few ($\sim 3$) binary orbits. The resulting
precession rates are shown as purple dots in \figref{fig:precession_tmep}, and
they lie on the same curve. For the simulations with varying $\alpha$-values,
the relation between the precession rate and the temperature is not as
pronounced, and simulations that start in the low eccentric disk state do not
fit the model anymore (two rightmost big red points in the plot).
In \figref{fig:peri_ecc}, it can be seen that the disks build up eccentricity
when the longitude of pericenter is shifting toward the binary's periastron and
reduce their eccentricity when the longitude of pericenter is shifting away from
the binary periastron position. This causes disks with lower precession rates to
have more time to increase their eccentricity and have therefore a larger
variance in eccentricity oscillations.
\subsection{Solid body precession}
Typically, in our simulations, the disks precess as a solid body, that is, the
radial profile of the longitude of pericenter is constant across the whole disk.
In some simulations, however, we observe deviations from the solid body
precession. These mostly appear as jumps in the longitude of pericenter, at
radii at which the opacity changes due to the temperature gradient in the disk.
In our opacity models, the variation with temperature is given by different
power-laws. Abrupt changes in these split the disk into regions with different
temperature gradients, which affects the propagation of the perturbations in the
disk as sound waves are partially reflected at these discontinuities. This can
cause the disk to split into two regions at the power-law transition that
precess independent of each other.

\begin{figure}
	\centering
	\includegraphics[width=88mm]{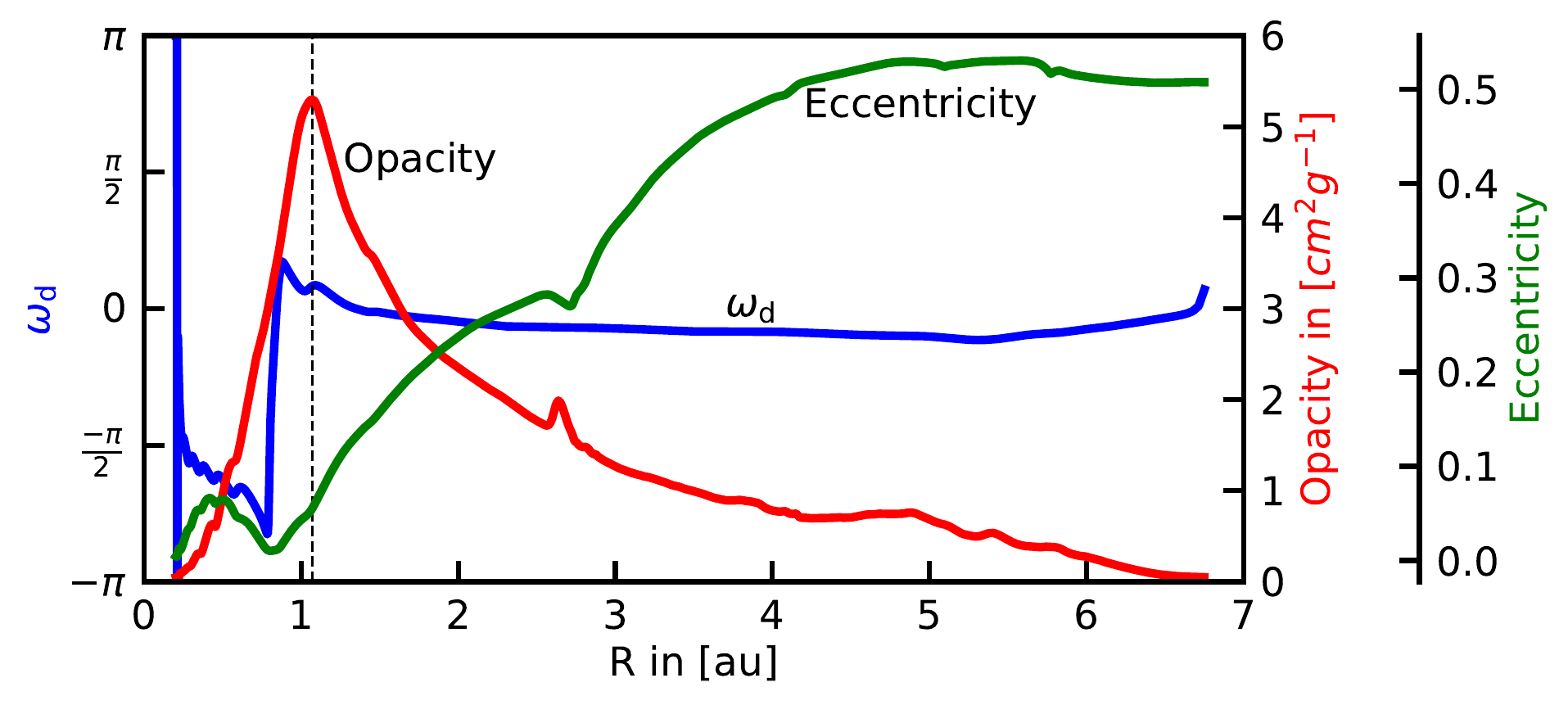}
	\caption{Snapshot at $T = 662.74\,T_\mathrm{bin}$ of the radial profile of
	the longitude of pericenter, opacity and eccentricity of the fiducial model.
	The dashed vertical black line indicates the position of the change in the
	opacity power-law at $T=1100\,\mathrm{K}$. The longitude of pericenter
	becomes undefined for $e = 0$ at the inner boundary.}
	\label{fig:whe_example}
\end{figure}

\begin{figure}
	\centering
	\includegraphics[width=88mm]{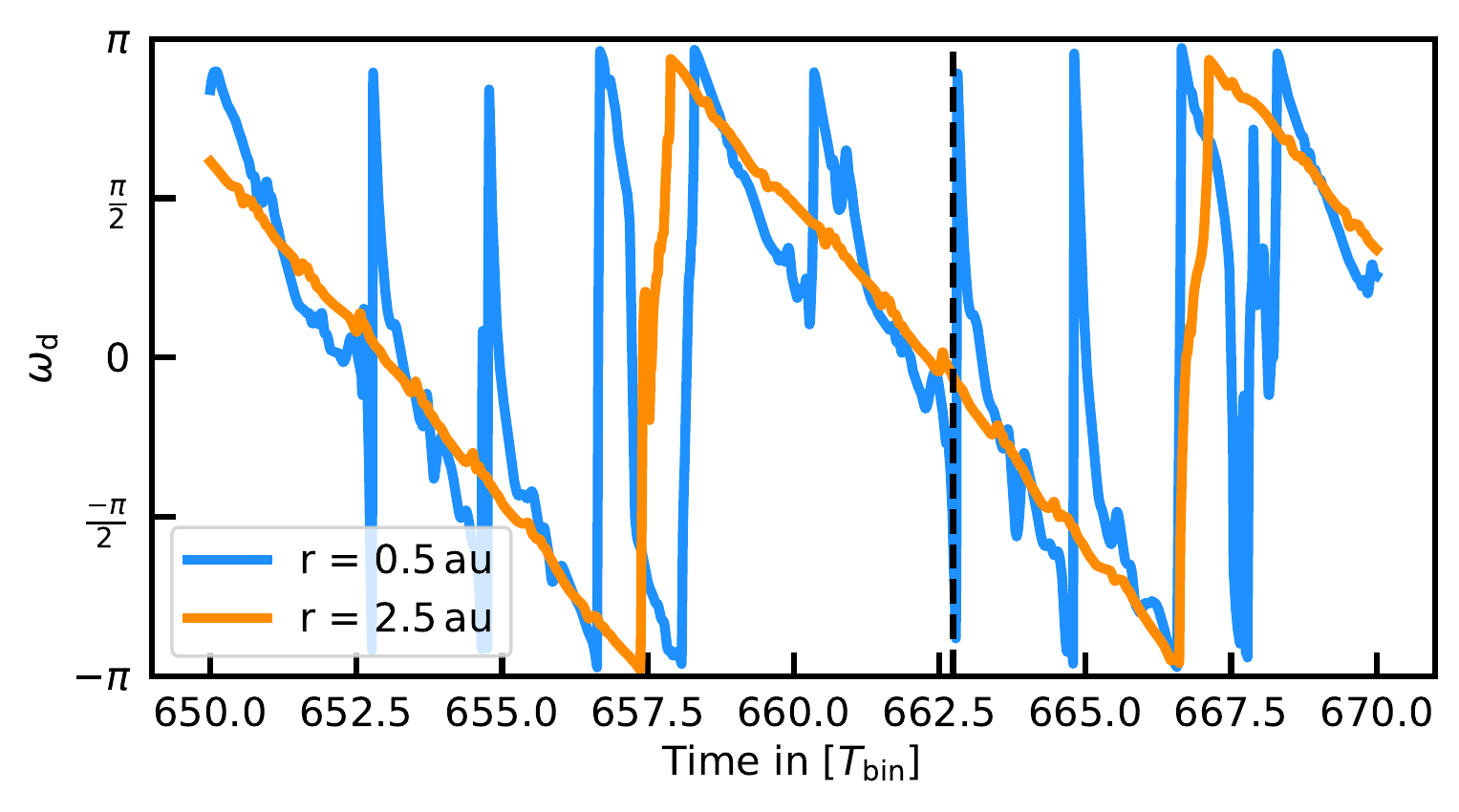}
	\caption{Time evolution of the azimuthally mass-weighted average of the
	 longitude of pericenter at two different radii for the fiducial model. The
	 dotted line indicates the time the snapshot used in
	 \figref{fig:whe_example}.}
	\label{fig:w_example}
\end{figure}

An example of this effect is displayed in \figref{fig:whe_example} which shows a
snapshot of the disk longitude of pericenter, opacity, and eccentricity of the
fiducial model just after the binary periastron passage when the spiral arms had
propagated to the inner parts of the disk. The change in wave propagation
induced by the opacity law splits the disk's longitude of pericenter at $ \sim
1\,\mathrm{au}$ where the temperature in the disk surpasses the dust sublimation
temperature of $T\approx1100\,\mathrm{K}$ and the temperature dependence of the
opacity changes from $\kappa\sim T^1$ to $\kappa \sim T^{-9}$.
\figref{fig:w_example} shows that the precession at the inner region of the disk
(blue line) becomes irregular and independent of the outer disk region (red
line). By the time the binary reaches apastron, the perturbations have
dissipated, and the inner region of the disk starts to realign to the outer
region of the disk. While the region's precession is disconnected, the
eccentricity of the disk is reduced at the boundary between the two regions,
lowering the eccentricity throughout the whole disk. For our low mass simulation
$M_\mathrm{disk} = 2\cdot10^{-3}\mathrm{M}_\odot$, the longitude of pericenter
breaking happens when the temperature surpasses $155\,\mathrm{K}$ and the
power-law changes from $\kappa \sim T^2$ to $\kappa \sim T^{-7}$. This
transition happens further outward at $2\,\mathrm{au}$, causing the eccentricity
damping of the differential precession to be more effective (compare the onset
of longitude of pericenter breaking for the low mass simulation (blue line) at
$t = 100\,T_\mathrm{bin}$ versus the onset of breaking for the fiducial model
(red line) at $t = 600\,T_\mathrm{bin}$ in \figref{fig:log_mass_ecc}).

\begin{figure}[!htb]
	\centering
	\includegraphics[width=88mm]{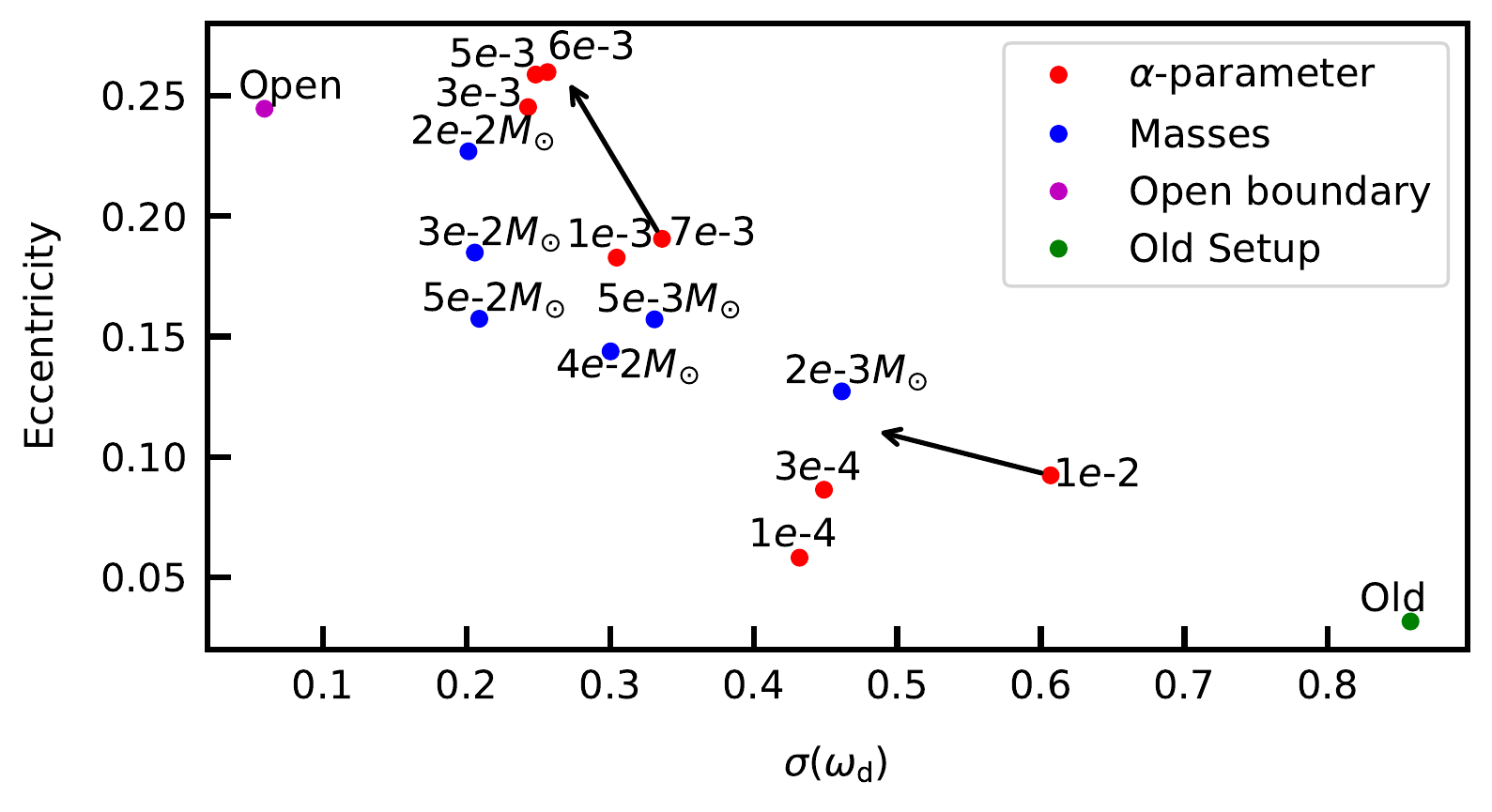}
	\caption{Disk eccentricity plotted against the standard deviation of the
	longitude of pericenter of all cells. The values mass averaged and time
	averaged over 200 snapshots at binary periastron from $T =
	1300\,T_\mathrm{bin}$ to $T = 1500\,T_\mathrm{bin}$. The old setup is using
	the standard parameters presented in \citet{muller2012circumstellar}. The
	arrows indicate the state after $2000\,T_\mathrm{bin}$.}
	\label{fig:EdW}
\end{figure}

We calculated the mass-weighted standard deviation of the longitude of
pericenter $\sigma\left(\omega_\mathrm{d}\right)$ for all cells in the disk and
found an anticorrelation between $\sigma\left(\omega_\mathrm{d}\right)$ and disk
eccentricity, see \figref{fig:EdW}. This effect acts on top of the other effects
that influence the disk eccentricity that we discussed in this work. For the
high mass simulations in \figref{fig:EdW}, the drop in eccentricity without an
increase in $\sigma\left(\omega_\mathrm{d}\right)$ is caused by radiative
damping. Simulations with a low $\alpha$-value develop a lower eccentricity than
other simulations with comparable $\sigma\left(\omega_\mathrm{d}\right)$,
indicating that the drop is due to their smaller size rather than differential
precession.

Disks that enter a quiet state do not precess and have a highly varying
longitude of pericenter; but due to their very low eccentricity, it is not clear
whether their high $\sigma\left(\omega_\mathrm{d}\right)$ is damping their
eccentricity or their low eccentricity makes determining the longitude of
pericenter less precise. Additionally, this effect depends on the inner boundary
condition. An open inner boundary condition removes mass and reduces the
temperature gradient across the disk, reducing the influence of power-law
changes in the opacity table. This again highlights the importance the inner
boundary has in our simulations.

\begin{figure}[!htb]
	\centering
	\includegraphics[width=88mm]{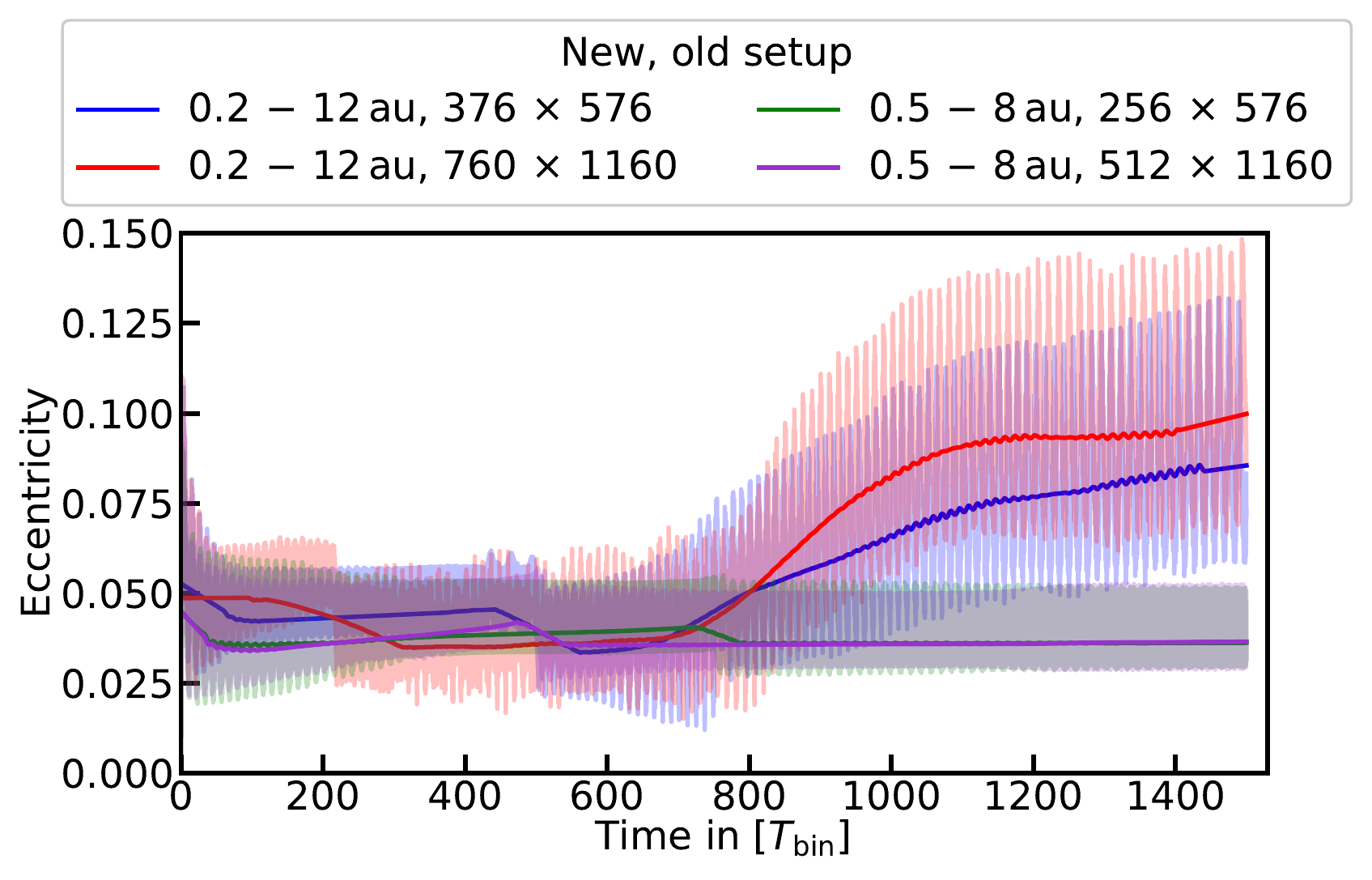}
	\caption{Time evolution of the mass-weighted disk eccentricity for the
    parameters of the standard case of \citet{muller2012circumstellar} that is
    shown in green, and the parameters given in \tabref{table:parameters} but
    with $\alpha = 10^{-2}$.}
	\label{fig:log_old_ecc}
\end{figure}

\section{Discussion}
\label{sec:discussion}
\subsection{Comparison to previous work}
\label{sec:previous}
\begin{figure}[!t]
	\centering
	\includegraphics[width=88mm]{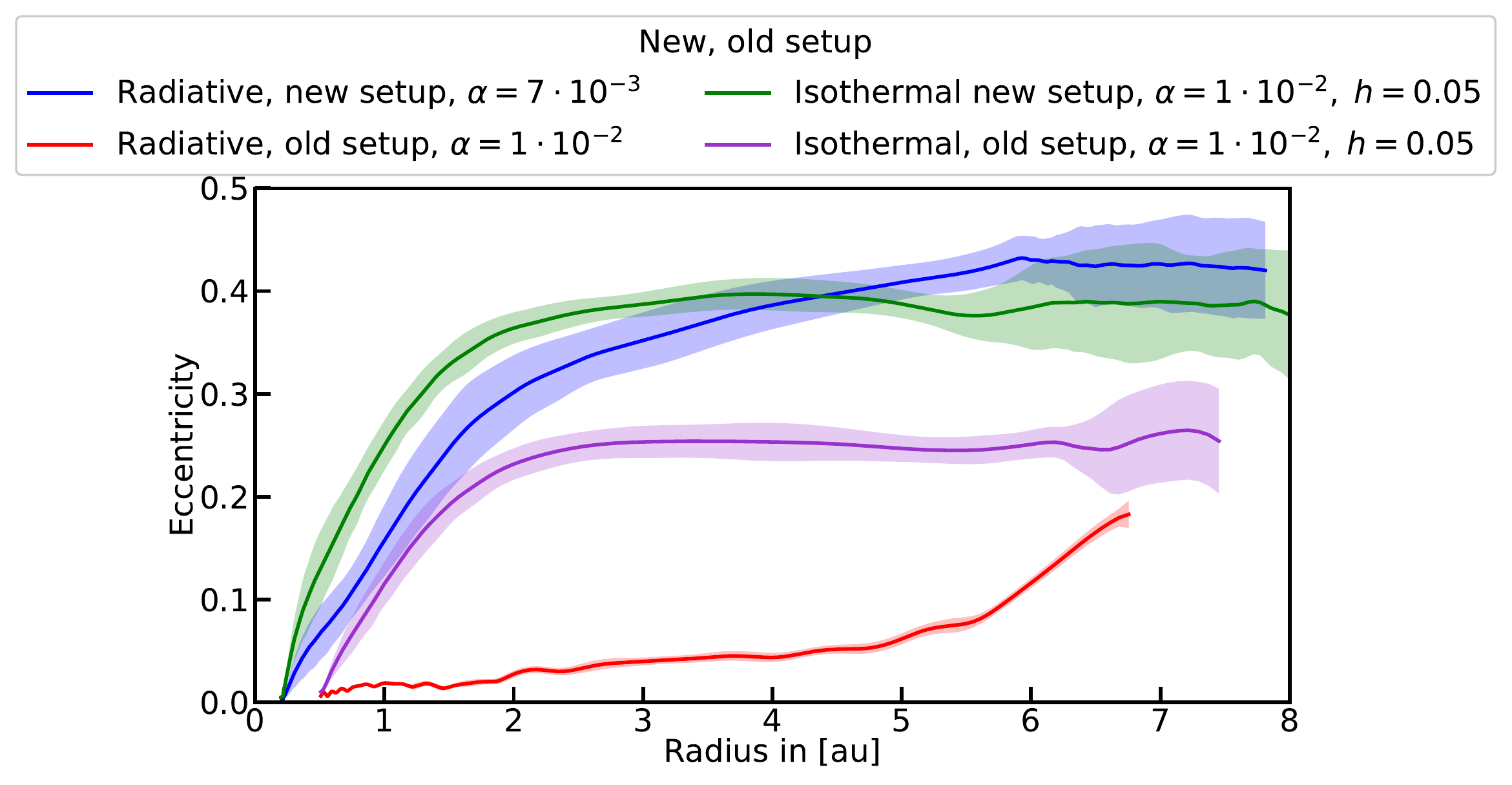}
	\caption{Radial profiles of the time averaged mass-weighted disk
    eccentricity for our new setup and the standard models from
    \citet{muller2012circumstellar}. Solid lines are averaged over 200 snapshots
    taken at binary apastron. The shaded areas show the $1\,\sigma$ variations.}
    \label{fig:time_avg_log_iso_old_ecc}
\end{figure}

Previous studies about radiative disks in the system of $\gamma$-Cephei
(\cite{muller2012circumstellar}, \cite{marzari2012eccentricity}) found only low
disk eccentricities of $e_\mathrm{d} \sim 0.05$, which is in contrast to our new
simulations presented here. We attribute the differences to a combination of the
high viscosity values used for their fiducial models of $\alpha \approx
10^{-2}$, a too low resolution, and a too-small simulation domain. For such a
high viscosity, the disks stay in a quiet state with low eccentricity and no
precession. This quiet state is robust to changes in numerical parameters,
giving a false sense that lower resolutions and smaller domains have converged.
\figref{fig:log_old_ecc} shows simulations for the standard parameters from
\cite{muller2012circumstellar} (green curve) and their convergence test (purple
line) and simulations with the domain size of our standard model
\tabref{table:parameters} (red and blue lines). After we increased the domain
size, the disks started to develop eccentricity and precession after $700\,
T_\mathrm{bin}$ of simulation time, which is longer than the simulation time of
the convergence tests in \cite{muller2012circumstellar}. The difference in disk
structure for the setups is shown in \figref{fig:time_avg_log_iso_old_ecc}. The
plot compares the radial eccentricity profile of radiative and locally
isothermal models for the new and old setup, averaged over 200 snapshots taken
at binary apastron. The curves for the old model (red and purple) use the
original data from \citet{muller2012circumstellar} with their standard viscosity
parameter of $\alpha = 10^{-2}$. The radiative simulation on the new setup uses
a lower viscosity of  $\alpha = 7\cdot10^{-3}$ because the $\alpha =
1\cdot10^{-2}$ simulation has not reached an equilibrium state at the end of the
simulation time of $2000\,T_\mathrm{bin}$. \figref{fig:time_avg_log_iso_old_ecc}
highlights the drastic difference between a quiet state and an excited state.
The eccentricity of the disk in the quiet state is very low ($e_\mathrm{d} \leq
0.05$), and it does not precess. This means that the disk is always in the same
state when the binary reaches apastron, as indicated by the vanishing standard
deviation of the eccentricity. On the other hand, the dynamics of an excited
disk also depend on the disk's own orientation inside the binary and oscillates
with the disk's precession rate. It can be seen that the dynamics of our
radiative simulations are closer to the dynamics of locally isothermal
simulations rather than to the radiative simulations performed in previous
studies. It is still not clear why radiative simulations perform worse than
locally isothermal simulations at low resolutions. Despite locally isothermal
and radiative simulations being similar at high resolutions, we recommend
performing fully radiative simulations as locally isothermal simulations are
overestimating the disk's eccentricity as they are missing effects like
radiative damping and breaking of the solid body precession that can damp
perturbations in the disk.
 
\subsection{Mass loss}
The mass-loss rates we found from mass ejection, are negligible compared to the
mass accretion rates we found in our simulation with an outflow inner boundary
condition. Allowing mass to leave through the inner boundary limits the disk's
lifetime in our simulations to $\lessapprox 10^5\,\mathrm{yr}$ for $\alpha$ =
$10^{-3}$. This represents a lower estimate for the disk's lifetime as the mass
loss is overestimated by our boundary condition. In realistic systems, the
circumprimary disk will be accompanied by a circumsecondary disk and a
circumbinary disk that all exchange mass between each other, as shown in
numerical simulations (e.g.,
\citet{2002A&A...387..550G,kaigorodov2010structure,nelson2016dynamics,munoz2016pulsed}).
The circumstellar disks could then be replenished with material from the
circumbinary disk, which increases their lifetime \citep{monin2007evolution}.
Observations of GG Tau A, which is a triple star system with a comparable
configuration to $\gamma$-Cephei that hosts a disk around the primary star and a
massive circumbinary disk of $\sim 0.15\, M_\odot$, find mass accretion rates
from the circumbinary disk onto the circumstellar disks of $\dot{M} \approx
6.4\cdot 10^{-8} M_\odot/\mathrm{yr}$ \citep{phuong2020ggtau}. The infalling
mass is expected to be accreted in an approximately equal ratio onto the
circumstellar disks \citep{dittman2021accretionratio}. Extrapolating the mass
loss rate from our open inner boundary condition simulation to the point it
would be balanced out by a mass accretion rate of $3.2\cdot 10^{-8}
M_\odot/\mathrm{yr}$ results in a disk mass of $M_\mathrm{d}\approx
8\cdot10^{-4}\,M_\odot$. Since our inner outflow boundary condition likely leads
to an artificially enhanced mass accretion, circumstellar disks with masses of a
few $10^{-3}\,M_\odot$ could have the same lifetime as the circumbinary disk
that feeds them. A similar mass has been estimated from observations for the
disk around GG Tau Aa \citep{dutrey2014ggtau}. This indicates that the
circumstellar disk lifetime inside binaries is not necessarily a constraint for
planet formation as they are fed by circumbinary disks that have been observed
to have similar lifetimes as disks around single stars
\citep{kuruwita2018multiplicity}

\subsection{Accretion shocks}
The mass infalling onto the circumprimary disk is accreted in pulses
\citep{munoz2016pulsed}, leading to shocks and shock heating in the disk
\citep{nelson2016dynamics} and could cause the disk to develop a third spiral
arm \citep{picogna2013three}. Since those studies did not resolve the
circumstellar disks at a high enough resolution or evolved the system long
enough to build up eccentricity, the effects of pulsed accretion on the disk
dynamics are poorly understood. We conclude from these studies that the
perturbations of the infalling material onto the circumprimary disk are minor
compared to the gravitational perturbations of the companion. Therefore, we
expect the trends we found to hold even if a mass accretion onto the disk is
added.

All our simulations are two-dimensional. Running the same setup with three
dimensions will affect the disk dynamics. Due to an extra degree of freedom for
the gas motion, shock heating is less effective in 3D than in 2D
\citet{bate2003heating3d}. As the companion excites strong shocks inside the
disk that produce significant heat, our 2D approach likely overestimates the
disk's temperatures. Additionally, density wave propagation and their deposition
of angular momentum into the disk could be different in three dimensions
\citet{lubow1998threedim, lee2015vertical} which could affect the disk dynamics.
While comparable 3D simulations did find differences to 2D simulations
\citet{picogna2013three,zhu2015_3d}, these were located at the disk's surface.
Thus, we expect our 2D simulations to be comparable to the dynamics found in a
3D simulation at the mid-plane, where most of the mass resides.

\section{Summary}
\label{sec:summary}
We revisited the $\gamma$-Cephei system by performing grid-based,
two-dimensional hydrodynamical simulations to study the dynamics of the disk
around the primary star in an eccentric binary star. All of our simulations
include viscous heating and radiative cooling. The disk dynamics are analyzed in
terms of eccentricity and precession rate. We carefully separated physical from
numerical behavior by investigating first the effects of numerical parameters
before looking into the impact of physical conditions such as disk mass and
viscosity.

We found that previous studies on radiative disks in close binaries used a too
low resolution and too small simulation domain to properly resolve the disk
dynamics. When using appropriate resolution and domain size, we observe
considerably higher disk eccentricities. To reach the higher eccentric state, a
resolution of at least six grid cells per scale height is required. Also, the
simulation domain should contain the whole Roche lobe of the primary at phases
of its orbit to avoid artificial mass loss through the outer boundary. The disk
eccentricity linearly declines with increasing inner domain radius. This could
imply that the disk becomes more circular as the inner region is cleared by
photoevaporation.

It was already noted in \citet{paardekooper2008planetesimal} that the numerical
methods used can change the outcome of the simulations. These effects were not
observed in later studies such as \citet{muller2012circumstellar} due to their
fiducial model being caught in a quiet state that is robust to numerical
changes.
This quiet state is characterized by a low eccentricity and no disk precession.
These occur in the initial phase of the simulations for viscosities larger than
$\alpha \geq 7\cdot10^{-3}$, and they can last for over $\approx
700\,T_\mathrm{bin}$. Eventually, the disks do become eccentric and start
precessing. We found all of our disks become excited across different numerical
and physical parameters as well as different codes (additionally to our
\textsc{Fargo} code, we tested the \textsc{Pluto} code). We are therefore
confident that the excited state is physical and that the very low
eccentricities reported for radiative disks in
\citet{muller2012circumstellar,marzari2012eccentricity} are inaccurate.

In agreement with \citet{muller2012circumstellar} and
\citet{marzari2012eccentricity}, we find that it is important to perform more
realistic simulations that include viscous heating and radiative cooling.
Locally isothermal models overestimate the disk's eccentricity because they are
missing effects like radiative damping or breaking of the solid body precession
of the disk which can dissipate perturbations and damp the disk's eccentricity.

We used reflective boundaries at the inner radius for our simulations so that
the disks can reach an equilibrium state. This boundary can physically be
interpreted as the disk extending to the star's surface. We also ran a test
simulation with an outflow inner boundary condition which coarsely resembles a
cavity with perfect accretion, possibly created by a stellar magnetosphere.
Despite these being two opposite extremes for the inner boundary condition, they
lead to a similar early development of the disks. The disk evolution diverges
over time as the open inner boundary simulation rapidly loses mass. We conclude
that the trends we find would also apply to a more realistic boundary condition.

For the mass-weighted and time-averaged disk eccentricity, we found values from
$0.06$ to $0.27$ (\figref{fig:ecc_vis}). The disk's size is given by a balance
of viscous versus gravitational torques. In our simulations, we found values
from $4\,\mathrm{au}$ for disks with low viscosities, which is identical to the
orbital stability limit for massless test particles \citep{holman1999long,
pichardo2005circumstellar}, up to a maximum disk size of $5.8\,\mathrm{au}$ for
higher viscosities (\figref{fig:vis_radius}). The disk size, or indirectly the
viscosity, is an important factor for the disk eccentricity. Smaller disks are
less affected by the perturbations of the secondary and become less eccentric.
The disks' eccentricity can also be reduced by radiative damping. In our
simulations, radiative damping becomes effective for hot disks with $h \geq
0.06$. The eccentricity of the disk oscillates with the same period as the
longitude of pericenter, and we found that the amplitude of the oscillation
increases with the precession period. Our simulations did not include the
effects of self-gravity, which can change the radial alignment of the longitude
of pericenter and can damp the disk's eccentricity
\citet{marzari2009eccentricity}.

In all our simulations, we observed a slow retrograde precession of the disk.
The precession rate depends linearly on the mean disk temperature, or
equivalently, quadratically on the aspect ratio, as given by
eq.\,(\ref{eqn:disk_precession}). This relation is in good agreement with the
values found in \cite{kley2008simulations}, who performed locally isothermal
simulations for disks in circular binaries in the context of the superhump
phenomenon in cataclysmic variables. From eq.\,(\ref{eqn:disk_precession}), we
can expect that disks with low temperature will change to a prograde precession
which might occur for disks with small disk masses during the dissipation phase.

We noticed in our simulations that the disk needs to precess as a solid body to
be able to develop high eccentricities. Changes in the opacity power-law can
break the longitude of pericenter alignment inside the disk, splitting the disk
into an inner and outer region that precess independently of each other,
reducing the overall disk eccentricity (see \figref{fig:whe_example}). We
correlate the disk eccentricity to the standard deviation of the longitude of
pericenter across all cells and find a downward trend in \figref{fig:EdW}.  
We suspect that the low disk eccentricities found in \citet{martin2020evolution}
could be explained similarly. In their isothermal SPH simulations, they observed
that particles at different radii have different precession rates (see their
Fig\,1). This could have prevented their disks from reaching higher
eccentricities.

Concerning the impact on the planet formation process, the temperatures in our
models with realistic masses $(M_\mathrm{d} \leq 10^{-2}M_\odot)$ are cold
enough for silicate to be present in a solid-state throughout most of the disk.
For our coldest model $M_\mathrm{d} = 2\cdot10^{-3}M_\odot$ the snowline
($160\,\mathrm{K}$) is located in the middle of the disk at $2\,\mathrm{au}$ and
the dust sublimation temperature ($1100\,\mathrm{K}$) is reached near the inner
boundary at $0.5\,\mathrm{au}$. Due to the low disk mass, the heating of the
shocks is also weak. The spiral wave heats the outer parts of the disk to
$200\,\mathrm{K}$ while the heating strength is reduced in parts where the disk
temperature is already higher than $200\,\mathrm{K}$ from viscous heating. While
the ices are repeatedly sublimated and recondensed by the spiral waves, silicate
dust particles are mostly unaffected by the shock heating from the companion.
However, we find high eccentricities and a retrograde precession for disks in
close binaries over a wide range of parameters. High gas eccentricities have
been found to increase the collision velocities between particles inside the
disk, which slows down dust coagulation \citep{zsom2011coagulation} and makes
planetesimal accretion less likely to succeed
\citep{paardekooper2008planetesimal}. Therefore, our results support previous
studies that found that the standard planet formation channel is unlikely to
succeed in the $\gamma$-Cephei system as it is currently observed.

Alternative formation scenarios have been proposed to explain the increasing
number of planets detected in close binary stars (see
\citet{schwarz2016catalogue} or Fig.\,1 in \citet{Thebault2019Planets} for a
complete list of systems).
Instead of in situ formation, the planets could also have formed inside a
circumbinary disk and then be scattered and captured on a circumstellar orbit
\citep{gong2018scattering}. Additionally, they could have formed in a less
hostile environment and end up in a close binary through a variety of different
star-star scattering scenarios (e.g., \citet{marzari2007configuration},
\citet{malmberg2007encounters}, \citet{fragione2019dynamical}). Nevertheless,
there are multi-planet systems observed in close binaries, such as the
Kepler-444 system \citep{lillo2014kepler444} which hosts five small, coplanar
planets around the primary star. It is unlikely that they formed via scattering
events suggesting that planet formation can indeed succeed in close binaries
after all \citep{dupuy2016multiplanet}.

{\small
\section*{Acknowledgments}
\font\testa=ec-lmr9 at 10pt We would like to thank Alexandros Ziampras for
helping to utilize the \textsc{Pluto} code and for helpful discussions about the
disk's thermodynamics. The  authors  acknowledge  support  by  the  High
Performance  and  Cloud Computing  Group at the Zentrum für Datenverarbeitung of
the University of Tübingen, the state of Baden-Württemberg through bwHPC, and
the German Research Foundation (DFG) through grant  INST 37/935-1 FUGG. All
plots in this paper were made with the Python library {\testa matplotlib}
\citet{hunter2007matplotlib}.}
\bibliographystyle{aa}
\bibliography{references}

\begin{appendix}
\section{Test of different codes}
\label{sec:codes}

\begin{figure}
	\centering
	\includegraphics[width=88mm]{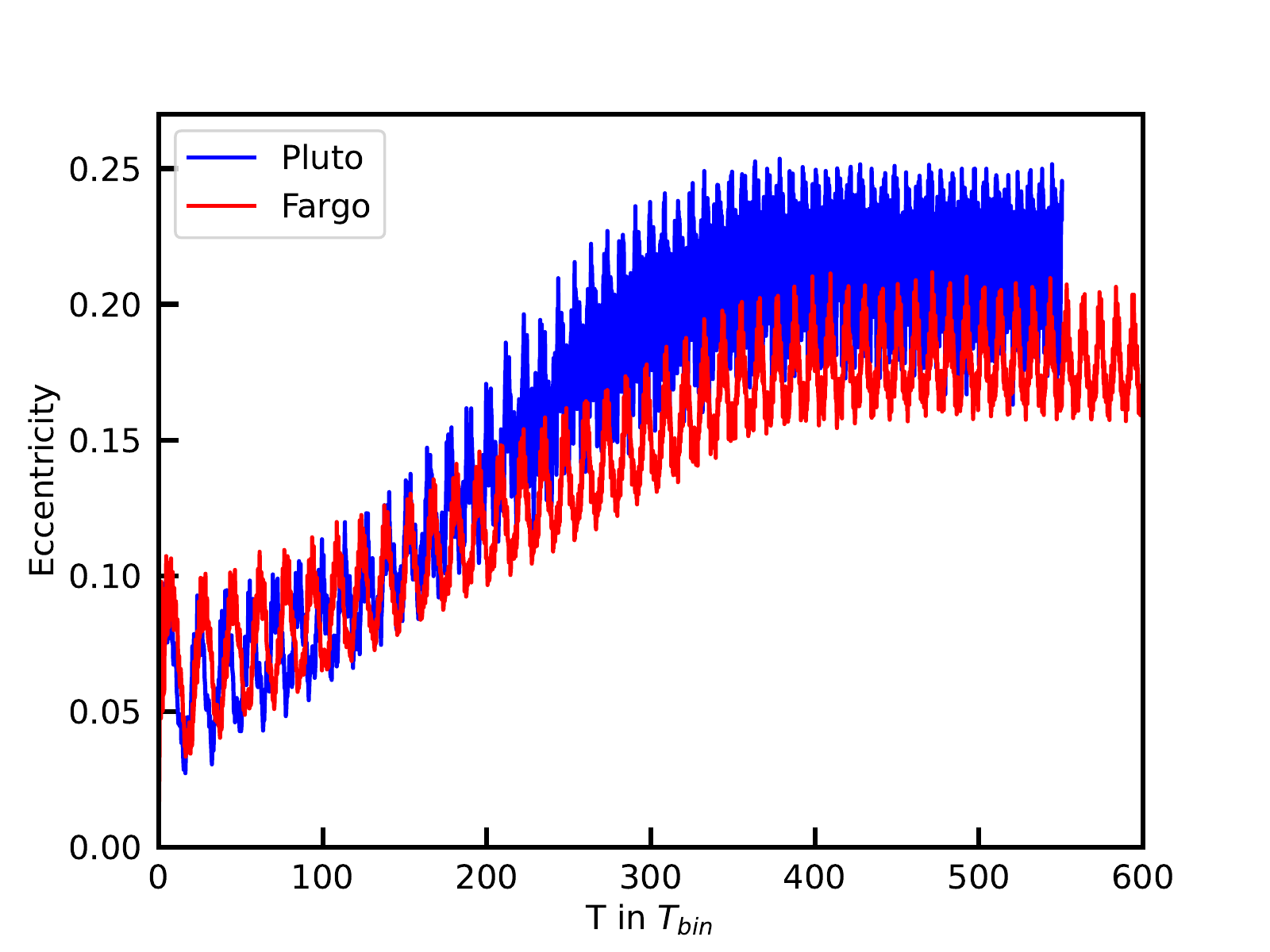}
	\caption{Time evolution of the disk eccentricity using the physical setup of
        the standard model. Displayed are the results of two different codes
        that differ slightly in their resolution and domain size, see text for
        details.}
	\label{fig:pluto_ecc}
\end{figure}

To corroborate and validate our findings, we performed supplementarily
simulations for the standard model using two different codes. Our results are
presented in Fig.\,\ref{fig:pluto_ecc}. All simulations use the physical setup
of the standard model. The \textsc{Fargo} run uses $760\times1160$ active cells
on a domain ranging from $0.2 - 12\,\mathrm{au}$. The \textsc{Pluto} run uses
the $648 \times 1160$ cells but has its domain size reduced from $0.3 -
10\,\mathrm{au}$ to produce the same cell size as in the \textsc{Fargo}
simulation. The inner boundary is larger to reduce computation time, and the
outer boundary is smaller for numerical stability. The \textsc{Fargo} simulation
utilizes the Fargo method to speed up the simulations while the \textsc{Pluto}
run does not apply the Fargo method and therefore requires approximately ten
times more timesteps due to the smaller $\Delta t$. The \textsc{Pluto} code also
uses a slightly different definition of the coefficient of the kinematic
viscosity of $\nu = \alpha H c_s$, which is a factor $\gamma^{-1/2}$ smaller
than in our \textsc{Fargo} code, where $\gamma$ is the adiabatic index.

For both codes, the disks enter an excited state with comparable eccentricity
evolution. The differences between the simulations can be explained primarily by
the different numerical methods used in the codes. The \textsc{Pluto} code
utilizes a Riemann-solver-based method that conserves total energy, at least in
the hydrodynamic part. In contrast, \textsc{Fargo} uses a second-order upwind
scheme that is not energy conserving and utilizes artificial viscosity to
stabilize discontinuities. This energy-conserving property makes shock heating
more effective and causes the \textsc{Pluto} simulation to develop a higher disk
temperature in the outer regions ($r > 1\,\mathrm{au})$. Due to the higher
temperature, \textsc{Pluto} produces a faster precessing disk which is in
agreement with our predictions. The \textsc{Pluto} simulation has a steeper
eccentricity increase from the inner boundary, which results in an overall
higher disk eccentricity. The overall similar dynamical evolution of the disk
for the different codes lends strong support to our conclusion that the
eccentric disk state is physical.

\end{appendix}
\end{document}